\documentclass[aps,pre,reprint,floatfix,longbibliography]{revtex4-1}
\usepackage{amsmath}
\usepackage{amssymb}
\usepackage{bm}
\usepackage{environ}
\usepackage{graphicx}
\usepackage{hyperref}
\usepackage{mathtools}

\newcommand{\code}[1]{\texttt{#1}}  % typewriter for computer code
\newcommand{\mx}[1]{\mathbf{#1}}    % boldface for vector and matrix variables
\newcommand{\soft}[1]{\textsc{#1}}  % slant for software package names 
\newcommand{\op}[1]{\mathcal{#1}}   % calligraphic for operators

\newcommand{\eqlabel}[1]{\label{eq:#1}}
\newcommand{\eq}[1]{\eqref{eq:#1}}
\NewEnviron{eqsplit}[1]{%
  \begin{align}\eqlabel{#1}
    \begin{split} 
      \BODY 
    \end{split}
  \end{align}%
}
\newcommand{\Fig}[1]{Figure~\ref{fig:#1}}
\newcommand{\fig}[1]{fig.~\ref{fig:#1}}
\newcommand{\figlabel}[1]{\label{fig:#1}}
\newcommand{\dblfigure}[3]{         % page wide figure
  \begin{figure*}[htbp]
    \centerline{#1}
    \caption[]{#2}
    \figlabel{#3}
  \end{figure*}%
}
\newcommand{\sglfigure}[3]{         % column wide figure
  \begin{figure}[htbp]
    \centerline{#1}
    \caption[]{#2}
    \figlabel{#3}
  \end{figure}%
}
\newcommand{\seclabel}[1]{\label{sec:#1}}
\newcommand{\secref}[1]{\ref{sec:#1}}
\newcommand{\subsecn}[1]{Subsection~\secref{#1}}

\renewcommand{\@}{\partial}         % partial differential
\newcommand{\ee}[1]{\mathrm{e}^{#1}}% exponential
\newcommand{\Mx}[1]{\begin{bmatrix}#1\end{bmatrix}}   % matrix literal

\newcommand{\ms}{\mathrm{ms}}       % millisecond

\newcommand{\Arpack}{\soft{Arpack}}
\newcommand{\Auto}{\soft{Auto}}

\newcommand{\Dedalus}{\soft{Dedalus}}

\newcommand{\0}{\mx{0}}             % zero vector or zero matrix
\providecommand{\abs}[1]{\left\lvert#1\right\rvert}     % abs value
\newsavebox{\Brabox}\newsavebox{\Ketbox}
\newcommand{\braket}[2]{            % bracket notation, correct sizing
  \savebox{\Brabox}{\ensuremath{\displaystyle{#1}}}%
  \savebox{\Ketbox}{\ensuremath{\displaystyle{#2}}}%
  \left\langle 
    \usebox{\Brabox}%
    \phantom{\makebox[0pt][l]{\usebox{\Brabox}\usebox{\Ketbox}}}%
    \,\right|\left.%
    \phantom{\makebox[0pt][l]{\usebox{\Brabox}\usebox{\Ketbox}}}%
    \usebox{\Ketbox}
  \right\rangle%
}    
\newcommand{\bra}[1]{\langle #1 \,|}    % halves of 
  \newcommand{\ket}[1]{|\, #1 \rangle}  %   a bracket
\newcommand{\conj}[1]{\overline{#1}}    % complex conjugation
\renewcommand{\d}{\mathrm{d}}       % ordinary differential
\newcommand{\diag}[1]{\mathrm{diag}\left(#1\right)} % diagonal matrix
\newcommand{\E}[1]{\times10^{#1}}   % decimal exponent
\newcommand{\Ft}[1]{\mathcal{F}\left[#1\right]}  % Fourier transform
\newcommand{\Fti}[1]{\mathcal{F}^{-1}\left[#1\right]}  % .., inverse
\renewcommand{\H}{^{\dagger}}       % Hermitian conjugate (transposed+cc)
\newcommand{\Heav}{\mathrm{H}}      % Heaviside step function
\newcommand{\hot}{\textrm{h.o.t.}}  % higher order terms
\newcommand{\ii}{\mathrm{i}}        % imaginary unit
\newcommand{\intinf}{\int\limits_{-\infty}^{\infty}} % integral over R
\newcommand{\kron}[1]{\delta_{#1}}  % Kronecker's delta
\newcommand{\Linf}{L^{\infty}}      % max norm space
\newcommand{\Ltwo}{L^2}             % int square norm space
\providecommand{\norm}[1]{\left\lVert#1\right\rVert}    % norm
\renewcommand{\O}[1]{\mathcal{O}\!\left(#1\right)} % asymptotic order
\renewcommand{\o}[1]{o\!\left(#1\right)} % asymptotic order
\renewcommand{\Re}{\mathrm{Re}}     % real part
\newcommand{\Real}{\mathbb{R}}      % set of reals
\newcommand{\T}{^{\top}}            % transposed

\newcommand{\+}[2]{\def#1{{#2}}}
\newcommand{\1}[2]{\def#1##1{{#2}}}
\newcommand{\2}[2]{\def#1##1##2{{#2}}}
\newcommand{\3}[2]{\newcommand{#1}[3]{{#2}}}
\+{\A}{A} \+{\B}{B}     % coefficients in $\Cmp3$
\1{\a}{a_{#1}}          % component of linearised solution in the basis of efuncs
\1{\Cmp}{\mx{\Phi}_{#1}} % compatibility kernel, vector
\1{\Cmpf}{\hat{\mx{\Phi}}_{#1}} % .., Fourier image
\2{\cmp}{\Phi_{#1,#2}}  % .., component
\+{\c}{c}               % some comoving frame speed
\+{\ct}{\tilde{\c}}     % solitary wave propagating wave speed
\+{\cf}{\check{\c}}     % fast propagating wave speed
\+{\cs}{\hat{\c}}       % slow propagating wave speed
\1{\cso}{\cs^{(#1)}}    % .., asymptotic order
\+{\D}{\mx{D}}          % diffusion matrix
\+{\DA}{\mathrm{DA}}    % De-aliasing factor
\+{\Den}{\mathcal{D}}   % denominator in the critical strength formula
\+{\DNS}{_{\mathrm{DNS}}}   % relating to direct numerical simulation
\+{\dist}{\psi}         % "distance to the rest state"
\+{\eb}{\mx{e}_1}       % unit vector, direction of stimulation
\+{\Es}{E_s}                % ODE Jacobian's stable espace
\+{\Eu}{E_u}                % ODE Jacobian's unstable espace
\+{\eps}{\epsilon}      % convenient asymptotics parameter, frac power of \gam
\1{\efr}{\mx{v}_{#1}}   % right eigenfunction, enumerated
\2{\efro}{\mx{v}^{(#1)}_{#2}}% .., asymptotic term, enumerated
\1{\efl}{\mx{w}_{#1}}   % left eigenfunction, enumerated
\1{\eflf}{\hat{\mx{w}}_{#1}} % .., Fourier image
\2{\eflo}{\mx{w}^{(#1)}_{#2}}% .., asymptotic term, enumerated
\1{\eval}{\sigma_{#1}}  % eigenvalue, enumerated
\2{\evalo}{\sigma^{(#1)}_{#2}} % .., asymptotic term, enumerated
\+{\etol}{\eps_\mathrm{TOL}} % "numerical tolerance"
\+{\ff}{f}              % a scalar function
\+{\fff}{\hat{f}}       % .., Fourier image
\+{\f}{\mx{f}}          % reaction rates vector
\+{\fjac}{\f'}          % .., jacobian
\1{\fc}{f_{#1}}         % reaction components and their derivatives (\fi reserved by TeX)
\+{\Gam}{\Gamma}        % time scale matrix
\+{\gam}{\gamma}        % activator/inhibitor time scale ratio
\+{\gamc}{\gam_c}       % .., critical value in FHN
\+{\hr}{\bar{\mx{h}}}   % perturbation of the resting state
\+{\hs}{\hat{\mx{h}}}   % perturbation of the slow wave
\+{\Iex}{J}             % "external current" for AUTO continuation
\+{\i}{i}               % eval/efun/component enumerator
\+{\ign}{\chi}          % ignition mode of the critical nucleus
\+{\Jac}{\op{J}}        % Jacobian in the AUTO ODE
\+{\j}{j}               % eval/efun enumerator
\1{\Ker}{\mx{K}_{#1}}   % critical condition kernel
\+{\k}{k}               % eval/efun enumerator
\+{\L}{\op{L}}          % linearization operator
\+{\La}{\L\H}           % .., adjoint
\1{\Lo}{\L_{#1}}        % .., asymptotic term
\+{\Lams}{\Lambda_s}    % ODE Jacobian's stable evalues
\+{\Lamu}{\Lambda_u}    % ODE Jacobian's stable evalues
\+{\Length}{L}          % length of numeric interval for BVP
\+{\length}{l}          % length of numeric interval for DNS
\+{\l}{\ell}            % heuristics enumerator
\+{\Modes}{M}           % number of Chebyshev modes in DNS
\+{\Ngrid}{N}           % "grid size"
\1{\Num}{\mathcal{N}_{#1}}  % numerator of the critical amplitude
\+{\nuc}{\phi}          % critical nucleus at gam=0
\+{\Ps}{P_s}            % ODE Jacobian's stable espace projector
\+{\Pu}{P_u}            % ODE Jacobian's unstable espace projecto
\1{\Q}{\mu_{#1}}        % cross-correlation of perturbation with shift selector
\+{\q}{q}               % degenerate dependence power index
\+{\qq}{q}              % wavenumbers in Fourier transforms
\+{\shift}{s}           % the phase shift of the critical solution
\+{\Time}{T}            % time duration of DNS
\+{\t}{t}               % time (indep var)
\+{\Us}{U_s}            % stimulus magnitude
\+{\Usc}{U_{sc}}        % .., critical value
\+{\Usl}{\underline{U}_s} % .., .., lower bound
\+{\Usu}{\overline{U}_s}  % .., .., lower bound
\+{\u}{\mx{u}}          % reacting fields vector
\1{\ui}{u_{#1}}         % .., component (italic)
\2{\uio}{u^{(#2)}_{#1}} % .., .., asymptotic order
\+{\ur}{\bar{\u}}       % resting state
\1{\uri}{\bar{u}_{#1}}  % .., component (italic)
\+{\ut}{\tilde{\u}}     % generic propagating wave
\+{\uf}{\check{\u}}     % fast propagating wave
\+{\us}{\hat{\u}}       % slow propagating wave
\1{\usi}{\hat{u}_{#1}}  % .., component (italic)
\+{\uone}{\ui1}         % reacting field first component
\1{\uoneo}{\uio1{#1}}   % .., asymptotic term
\+{\uoner}{\bar{u}_1}   % .., resting state value
\+{\utwo}{\ui2}         % reacting field second component
\1{\utwoo}{\uio2{#1}}   % .., asymptotic term
\+{\utwor}{\bar{u}_2}   % .., resting state value
\+{\Vci}{\hat{V}_1}     % the critical solution, v-component, 1st order, nondim version
\2{\vi}{v_{#1,#2}}      % right efun component, enumerated. #1: ordinal number #2: component
\3{\vio}{v^{(#1)}_{#2,#3}}  % .., asymptotic term: #1: as order, #2: ordinal number, #3: component
\+{\w}{w}               % scalar unknown in the equation for left efun
\2{\wi}{w_{#1,#2}}      % left efun component, enumerated
\3{\wio}{w^{(#1)}_{#2,#3}}  % .., asymptotic term
\1{\X}{\bar{X}^{#1}}    % stimulus profile, scalar/component
\+{\Xb}{\bar{\mx{X}}}   % .., vector
\+{\Xbf}{\hat{\bar{\mx{X}}}}   % .., .., Fourier image
\+{\x}{x}               % space var
\+{\xf}{\xi}            % .., in moving FoR
\+{\xo}{\xi_0}          % .., coord of the pulse maximum
\+{\xp}{\xi'}           % .., dummy integration variable
\+{\xs}{\x_s}           % stimulus extent

\+{\fa}{\alpha}
\+{\fb}{\beta}
\+{\fbc}{\beta_c}      % critical value of beta

\+{\kbet}{\beta}
\+{\ktone}{\tau_1}
\+{\kttwo}{\tau_2}
\+{\kua}{u_a}
\+{\kud}{u_d}
\+{\kus}{u^*}
\+{\kM}{M}

\+{\mk}{k}
\+{\mtc}{\tau_c}
\+{\mti}{\tau_i}
\+{\mto}{\tau_o}
\+{\mtu}{\tau_u}
\+{\mth}{\theta}
\+{\mug}{u_g}
\+{\toyphi}{\varphi}    % dependent var in the toy model of $\cmp31$
\+{\toygam}{a}          % independent var in the toy model of $\cmp31$

\1{\fd}{\varphi_{#1}}   % first kinetic, or its derivative 
\+{\ud}{\omega}         % intermediate var
\+{\pwu}{\nu_1}         % power indices
\+{\pwv}{\nu_2}         %   in the degenerate
\+{\pwc}{\nu_c}         %   asymptotics

\begin{document}

\title{Predicting critical ignition in slow-fast excitable models} 

\author{Christopher D. Marcotte}
\affiliation{EPSRC Centre for Predictive Modelling in Healthcare, University
  of Exeter, EX4 4QJ, UK}

\author{Vadim N. Biktashev}
\affiliation{Department of Mathematics, University of Exeter, EX4 4QF, UK}

\date{\today}

\begin{abstract}
  Linearization around unstable travelling waves in excitable systems can be
  used to approximate strength-extent curves in the problem of initiation of
  excitation waves for a family of spatially confined perturbations to the
  rest state.  This theory relies on the knowledge of the unstable travelling
  wave solution as well as the leading left and right eigenfunctions of its
  linearization.  We investigate the asymptotics of these ingredients, and
  utility of the resulting approximations of the strength-extent curves, in
  the slow-fast limit in two-component excitable systems of FitzHugh-Nagumo
  type, and test those on four illustrative models.  Of these, two are with
  degenerate dependence of the fast kinetic on the slow variable, a feature
  which is motivated by a particular model found in the literature.  In both
  cases, the unstable travelling wave solution converges to a stationary
  ``critical nucleus'' of the corresponding one-component fast subsystem.  We
  observe that in the full system, the asymptotics of the left and right
  eigenspaces are distinct.  In particular, the slow component of the left
  eigenfunction corresponding to the translational symmetry does not become
  negligible in the asymptotic limit.  This has a significant detrimental
  effect on the critical curve predictions.  The theory as formulated
  previously uses an heuristic to address a difficulty related to the
  translational invariance.  We describe two alternatives to that heuristic,
  which do not use the misbehaving eigenfunction component.  These new
  heuristics show much better predictive properties, including in the
  asymptotic limit, in all four examples.
\end{abstract}


\maketitle

\section{Introduction}

Excitable systems represent a distinct modeling legacy in the context of
biological systems, especially regarding cardiac and neuronal cells.  These
models utilize the existence of a typically state-dependent threshold to
distinguish between relaxation and transiently amplified dynamics.  In classic
FitzHugh-Nagumo type models this threshold is explicitly the unstable branch
of the fast variable nullcline, volumes have been dedicated to investigations
around the response of excitable models to driving.

In the nomenclature of previous efforts~\cite{Bezekci-etal-2015}, we focus the
formalism of a ``stimulation by voltage", which corresponds to an initial
value problem for reaction diffusion models of the form,
\begin{equation}\eqlabel{rde}
  \@_\t \u = \D \@^2_\x \u + \Gam \f(\u),
\end{equation}
for a two-component field, $\u = [\ui1,\ui2]$. We assume
  that the model is non-dimensionalized with respect to the dynamic
  variables $\ui1$, $\ui2$, as well as the independent variables $\x$
  and $\t$. We restrict ourselves to the models where only the first
  component is diffusive, so $\D = \diag{1,0}$.  The time-scale
  separation of the dynamics of $\ui1$ and $\ui2$ is controlled by
  parameter $\gam$, with the time-scale of $\ui1$ considered fixed, so
  $\Gam = \mathrm{diag}(1,\gam)$. Further, we assume that
  $\f(\ur) = \0$ for a unique, asymptotically stable rest-state
  $\ur$. The limit $\gam\to 0$ designates the transition from moving
solutions ($\gam> 0$, $\c \neq 0$) to stationary solutions
($\gam = 0$, $\c=0$) of \eq{rde}.  In this work we concern ourselves
primarily with the transient dynamics in the vicinity of the traveling
wave solutions of these slow-fast systems.

Traveling wave solutions of \eq{rde} satisfy a nonlinear eigenvalue
problem posed on the real line $\x\in \Real$,
\begin{equation}\eqlabel{bvp}
  \0 = \D\@^2_\x \ut + \ct \@_\x\ut + \f(\ut),
\end{equation}
for the wave solution $\ut$ and the associated wave speed $\ct \neq 0$.
Additionally, we require that the wave approach the rest state asymptotically,
$\ut(\x\to\pm\infty) \to \ur$ so that the solution is localized in space, and a
homoclinic connection to and from the rest state in the co-moving frame.  The
simplest solutions to \eq{bvp} are two single-pulse waves: a faster,
stable wave $\uf$, and the slower, unstable wave $\us$, with $|\cs| < |\cf|$.
When speaking about general solutions of \eq{bvp}, we shall refer to
generic waves $\ut$ and associated generic speeds $\ct$.

The linear stability of these waves is determined by the eigenspectrum
$(\eval\i, \efr\i)$ of the operator $\L$, the linearization about
\eq{bvp},
\begin{equation}\eqlabel{linop}
  \L = \D \@^2_\x + \ct \@_\x + \fjac(\ut) ,
\end{equation}
which is guaranteed to have a marginal eigenfunction, $\L\efr{} = 0\efr{}$,
where $\efr{} = \@_\x \ut$, due to the translational symmetry.
Further, for the unstable asymptotic wave
solution $\us$, the operator $\L$ must have exactly one unstable mode,
$\eval{} > 0$, whose shape describes the fastest-growing mode in the
co-moving frame. We enumerate the modes in the
  decreasing order of the real parts, so for the unstable wave,
  \[
    \eval1 > \eval2 = 0 > \Re\left( \eval3 \right) \ge \dots .
  \]
Similarly, the inner-product over the domain,
\[
  \braket{ \efl{} }{ \L \efr{} } = \intinf \efl{}\H \L \efr{} \,\ \d\x,
\]
defines the set of adjoint eigenfunctions $\efl\j$ with eigenvalues
$\conj{\eval\j}$ satisfying the biorthogonality condition
$(\eval\j-\eval\i)\braket{\efl\j}{\efr\i} = (\eval\j-\eval\i)\kron{\i\j}$.
Equivalently, we may consider the right eigenfunctions of the adjoint linear
operator, $\La$,
\[
  \La(\u)\efl{} = \D\T  \@^2_\x \efl{} - \ct \@_\x \efl{} + \left[\fjac(\ut)\right]\T \efl{},
\]
where $(\cdot)\T$ stands for transposed and  $(\cdot)\H$ for Hermitian conjugate.

The transient dynamics in the vicinity of the unstable solution $\us$ is
important phenomenologically for understanding the initiation of e.g.,
electrical waves in heart tissue.  Ref.~\cite{Bezekci-etal-2015} used the properties
of the unstable wave to predict the minimal perturbations to the rest state
which lead to the generation of new excitation waves, using a particular
notion of `smallness' for the constructed perturbations which relies on
assumptions about the behavior of the solution $\us$ and its linearization in
the limit $\gam\to 0$.  In this work, we investigate the limitations of those
assumptions, and test alternative methods for constructing those minimal
perturbations.  We find that in several cases, the applicability of the
existing theory to traveling wave solutions of simple slow-fast models
predicts at the very least sub-optimal or occasionally unrealistically large
minimal perturbations.  We have in mind three specific examples of the models:
the classical FitzHugh-Nagumo model in the original
formulation~\cite{FitzHugh-1961} (FHN), the two-variable Karma 1994
model~\cite{Karma-1994} (Karma) and two-variable reduction of the Fenton-Karma
model due to Mitchell and Schaeffer~\cite{Mitchell-Schaeffer-2003} (MS). To our surprise
we found that Karma model is different from the other two in its asymptotics
in $\gam$, which has proved to be due to a specific form of dependence of the
fast kinetics $\fc1$ on the slow variable $\ui2$, namely via $\ui2^\kM$, with
$\kM>1$, whereas in the resting state $\uri2=0$.  To illustrate further the
specifics of such dependence, we have also considered a variation of the
FitzHugh-Nagumo model which also has this feature, that is, $\fc1$ depends on
$\ui2^3$; we shall call it ``FHN with cubic recovery'', or FHNCR for short.

The outline of this paper is as follows.  First we review the basics of sub-
and super-threshold response for classic excitable models of FitzHugh-Nagumo
type, the complication from embedding them in spatially extended media, and
how to distinguish sub- and super-threshold excitations in space.  Second, we
review the essential ingredients of the linear theory of critical excitation,
in the context of the asymptotics in the slow-fast time scale separation
between the activator and inhibitor subsystems.  We show that the approach
applied to slow-fast systems relies on misleading assumptions about the
asymptotic structure of the leading left and right eigenspaces and
particularly the form of the adjoint eigenfunctions.  We detail the
computation of the slow wave solutions, their eigenspectra, and the solution
of transient trajectories by direct numerical simulation.  Third, we propose
heuristics based on minimization principles which do not rely on the slow mode
corresponding to translational symmetry, and which outperform the previously
suggested skew-product motivated heuristic.  Finally we reinforce this
conclusion with numerical examples using several slow-fast systems, and use
these results to infer properties of the leading eigenspace and the effect of
degenerate nonlinearities on observing other nearby saddle solutions.

\section{Theory}

The mathematical problem is posed in the following way.
Given an initial condition $\u(0,\x) = \ur + \hr(\x; \xs, \Us)$ parameterized
by the properties $(\xs,\Us)$ of the stimulus, i.e. a perturbation to the rest
state,
determine
for which
configurations will the initial conditions eventually recruit the entire
domain -- excite the medium -- and for which configurations will it return
directly to the rest state.  In the language of coherent structures, this
corresponds to identifying the boundary of the basin of attraction isolating
the stable wave solution from the uniform rest state, and projecting this
infinite-dimensional manifold down to the shape-modifying parameter space of
$(\xs,\Us)$.  Throughout this work we shall use a parameterized perturbation
to the rest state,
\begin{eqsplit}{pert}
  & \hr(\x; \xs, \Us) = \Us \Xb{}(\x; \xs), \\
  & \Xb{}(\x; \xs) = \eb \X{}(\x; \xs), \\
  & \X{}(\x; \xs) = \Heav(\xs/2 - \x) \Heav(\x + \xs/2),
\end{eqsplit}
where $\eb=\Mx{1,0}\T$ and $\Heav(\x)$ is the Heaviside distribution,
so that $\norm{\hr(\x; \xs, \Us)}_{\infty} \equiv \Us$ and
\mbox{$\norm{\hr(\x; \xs, \Us)}_1 \equiv \xs \Us $}. %
The choice of $\eb=\Mx{1,0}\T$ means that we restrict
  consideration exclusively to perturbation of the first component of
  the system. This is in line with the prospective application of the
  theory to models of heart or nerve excitability, with the first
  component representing the transmembrane voltage, and the
  stimulation effected by external electric fields. Of course, in
  different application areas, different modalities of the stimulus may
  be more appropriate.  

\subsection{Linear theory of critical excitations}

Here we present a brief motivation for the linear theory of critical
excitations, and recount the assumptions of the method.  Given the
initial state $\u(0,\x) = \us(\x) + \hs(\x)$, where $\hs(\x)$ is
understood to be perturbatively small, then the dynamics of the state
subject to \eq{rde} can be understood through the linearization about
$\us(\xf)$, with $\xf = \x - \cs\t - \shift$ the co-moving frame
coordinate.  The presence of the shift parameter $\shift$
  here is due to translational invariance of the problem and its
  significance and issues associated with its choice are discussed
  below in~\subsecn{shift-selection}.  Expressing the linearized
dynamics in terms of the spectral expansion,
\[
  \L = \sum_{\k=1}^{\infty} \ket{\efr\k} \eval\k \bra{\efl\k },
\]
\[
  \ut(\t,\xf) = \us(\xf) + \sum_{\k=1}^{\infty} \a\k(\t) \efr\k(\xf),
\]
and recalling that $\Re(\eval{1}) > 0$, then we can consider the requirement
that the sole unstable mode is not excited due to the perturbation,
\begin{equation}\eqlabel{odesol}
  0 = \a1(\t) = \exp(\eval1 \t) \braket{ \efl1 }{ \hs(\xf) },
\end{equation}
where we identify the modal amplitudes of the linearization,
$\a\i(0) = \braket{ \efl\i }{ \hs(\xf) }$.  At time $\t=0$, we express the
initial perturbation $\hs$ to the critical solution $\us$ in terms of a
perturbation $\hr$ to the rest state $\ur$ so that $\xf + \shift = \x$,
\[
  \hs(\xf+\shift; \xs, \Us) = \hr(\xf+\shift; \xs, \Us) + \ur - \us(\xf),
\]
where the invariance of the rest state with respect to translational symmetry
manifests as a freedom in the origin of the rest-state perturbation, the shift
parameter $\shift$.  The freedom in choosing this origin must be dealt with
and the simplest functional method relies on computing the root of a scalar
function, whose form is heuristically determined based on assumptions about
the asymptotic structure of the eigenfunctions.  The combined system is,
\begin{eqsplit}{linsys}
  \Num{0} &= \Us \braket{ \Ker{0}(\xf) }{ \Xb(\xf+\shift; \xs) }, \\
  \Num{\l} &= \Us \braket{ \Ker{\l}(\xf) }{ \Xb(\xf+\shift; \xs) },
\end{eqsplit}
to be solved for $\Us$ and $\shift$ as a self-consistent system, for each
chosen $\xs$, with functional $\Ker{\l}$ constructed heuristically for
$\l=1,2,3$, in the next section.

\subsection{Shift selection}\seclabel{shift-selection}

Here we describe three heuristic arguments which lead to different forms of
$\Ker\l$, resulting in three different values of the shift $\shift$, and
ultimately three different predictions for the critical amplitude, $\Us$, per
chosen extent of the perturbation $\xs$.

The first heuristic seeks to minimize the amplitude $\Us$ chosen across all
the possible choices of the shift, $\shift$.  Defining $\Xb(\xf; \xs)$
according to \eq{pert}, so that $\Xb(\xf; \xs)$ is normalized in the
$\Linf$-norm, rearranging \eq{linsys} for the amplitude of the
perturbation,
\begin{equation}\eqlabel{Us}
  \Us = \braket{ \efl1(\xf) }{ \us(\xf) - \ur } \big/ \braket{ \efl1(\xf) }{ \Xb(\xf+\shift; \xs) },
\end{equation}
leads to the maximization of the denominator (as the numerator is independent of $\shift$), and the resulting condition,
\[
  \@_\shift \braket{ \efl1(\xf) }{ \Xb(\xf+\shift; \xs) } = 0,
\]
with $\@^2_\shift \braket{ \efl1(\xf) }{ \Xb(\xf+\shift; \xs) } < 0$.
Computing the derivative of this function with respect to $\shift$ can be
simplified using the definition of the inner product, which yields
\[
  \braket{ \efl1'(\xf) }{ \Xb(\xf+\shift; \xs) } = 0,
\]
so that we must find only the roots of a simple scalar equation, which when
combined with \eq{linsys} creates an appropriate solvability condition
for $\Us$.

The second heuristic assumes that the limiting feature of the linear theory is
the magnitude of the perturbation to the critical wave,
$\hs(\xf+\shift; \xs, \Us)$, and a minimization of the perturbation norm in
the $\Ltwo$ sense, with respect to the shift, ensures the dynamics are
appropriately linear and chooses the appropriate origin $\shift$.  Expressing
the perturbation to the wave in terms of the perturbation to the rest state
and minimizing with respect to $\shift$,
\[
  0 = \@_\shift \intinf \left(\hs(\xf+\shift; \xs, \Us)\right)^2 \, \d\x 
\]
subject to $\@^2_\shift \langle \dots \rangle > 0$, ultimately leads to
the condition
\[
  0 = \braket{ \efr2(\xf) }{ \hr(\xf+\shift; \xs, \Us) + \ur - \us(\xf) }.
\]
This reaffirms the importance of the Goldstone mode, and as with heuristic
$1$, the condition is combined with \eq{linsys} to create an appropriate
solvability condition for $\Us$.

The third heuristic requires that $\a2(0) = 0$, 
\[
  \a2(0) = \braket{ \efl2(\xf) }{ \hr(\xf+\shift; \xs, \Us) + \ur - \us(\xf) },
\]
following the same formulation as \eq{odesol}, and is equivalent to the
form used in Ref.~\cite{Bezekci-etal-2015}.  As with heuristic $2$, the condition is
then combined with \eq{linsys} to create an appropriate solvability
condition for $\Us$.

The values of $\Ker{l}(\xf)$ and $\Num{l}$ are summarized,
\begin{align*}
  \Ker{0}(\xf) &= \efl1(\xf), & \Num{0} &= \braket{ \efl1(\xf) }{ \us(\xf)-\ur }, \\
  \Ker{1}(\xf) &= \efl1'(\xf), & \Num{1} &= 0, \\
  \Ker{2}(\xf) &= \efr2(\xf), & \Num{2} &= \braket{ \efr2(\xf) }{ \us(\xf)-\ur }, \\
  \Ker{3}(\xf) &= \efl2(\xf), & \Num{3} &= \braket{ \efl2(\xf) }{ \us(\xf)-\ur }.
\end{align*}
Returning to \eq{linsys}, the solvability condition for $\Us$ is given by 
\begin{equation} \eqlabel{shifters-work}
  \braket{ \Cmp\l(\xf) }{ \Xb(\xf + \shift) } = 0           
\end{equation}
where 
\begin{equation} \eqlabel{shifters-struct}
  \Cmp\l(\xf) = \Num{0} \Ker{\l}(\xf) - \Num{\l} \Ker{0}(\xf),
\end{equation}
which determines the value of the shift $\shift$ for a given choice of $\l$.
We shall refer to the functions $\Cmp\l$ as ``shift selectors''.  The
determination of $\shift$ defines the solution of \eq{linsys} for $\Us$.
Note that according to \eq{shifters-work}, scaling $\Cmp\l$ by a nonzero
constant factor, or even a factor that is a function which is finite and
nonzero everywhere, does not change the answer. In the subsequent we shall
silently use this property to simplify the expressions where convenient.

Here we list the explicit forms of the shift selectors dropping dependence on $\xf$,
\begin{eqsplit}{Phil}
  \bra{\Cmp1} &= \braket{\efl1}{\us-\ur} \bra{\efl1'} \; \propto \bra{\efl1'},\\
  \bra{\Cmp2} &= \braket{\efl1}{\us-\ur} \bra{\efr2} - \braket{\efr2}{\us-\ur} \bra{\efl1},\\
  \bra{\Cmp3} &= \braket{\efl1}{\us-\ur} \bra{\efl2} - \braket{\efl2}{\us-\ur} \bra{\efl1}.
\end{eqsplit}
In the following we will use each heuristic-based shift selectors to predict
the critical excitation curve for several slow-fast models.

Note that since $\efr2=\@_x\us$, the components required for computing these
shift selectors, in addition to the critical pulse solution $\us$, are $\efl1$
and $\efl2$. Also, the stucture of \eq{Phil} means that $\efr2$ and
$\efl{1,2}$ are required up to a nonzero constant scaling factor.

The theory outlined so far is for generic
  excitable systems. In the subsequent, we look at the specifics of
  two-component systems with the fast-slow asymptotic structure.

\subsection{Asymptotic structure for the generic case}

The construction of the shift selectors $\Cmp\l$ requires knowledge of the
eigenfunctions, specifically $\efr2$, $\efl1$ and $\efl2$, and in view of
importance of the small parameter $\gam$ in many applications of the FHN
system, here we look at the the limit of small $\eps$, defined as
$\eps^2=\gam$.  When $\eps=0$, we have $\utwo(\x,\t)\equiv\utwor$, the
``voltage component'' of the critical solution is the stationary ``critical
nucleus'', with $\cs=0$ and $\usi1(\xf)=\nuc(\xf)$, such that
\[
  \nuc'' + \fc1(\nuc,\utwor) = 0,
\]
and the unstable mode $\ign(\xf)$ of the corresponding one-component
linearized problem defined by
\[
  \ign'' + \fc{11}(\nuc(\xf),\utwor) \ign = \eval1 \ign, 
  \qquad 
  \eval1>0. 
\]
Note for the future that both $\nuc(\xf)$ and $\ign(\xf)$ can be chosen as
even functions.  A naive expectation would be that for $\eps\to0$, we should
have $\lim\cs=0$, $\lim\us(\xf)=\Mx{\nuc(\xf),\utwor}\T$,
$\lim\efr1(\xf)=\lim\efl1(\xf)=\Mx{\ign(\xf),0}\T$, and
$\lim\efr2(\xf)=\lim\efl2(\xf)=\Mx{\nuc'(\xf),0}\T$.  These in fact were the
underlying assumptions in~\cite{Bezekci-etal-2015}.  In this section, we will
investigate the small-$\eps$ regime perturbatively to test these assumptions.

We look for the nonlinear wave solution as an expansion in $\eps$,
\[
  \us(\xf,\eps) = 
  \Mx{ \uoneo0(\xf) \\ \utwor}
  + \eps \Mx{ \uoneo1(\xf) \\ \utwoo1(\xf) }
  + \O{\eps^2},
\]
with the wave speed $\cs = 0 + \eps \cso1 + \O{\eps^2}$, and $\xf = \x-\cs\t$.
Substituting into the traveling wave equation \eq{bvp} and expanding in
$\eps$, we have in $\O{\eps^0}$,
\begin{align}
  & \uoneo0'' + \fc1(\uoneo0,\utwor) = 0, \eqlabel{u1ep0}\\
  & \utwo = \utwor ,
\end{align}
so we have $\uoneo0=\nuc$ and the naive assumption is true. In $\O{\eps^1}$,
\begin{align}
  & \uoneo1'' + \cso1 \uoneo0' + \fc{11}(\xf) \uoneo1 + \fc{12}(\xf) \utwoo1 = 0, \eqlabel{u1ep1}\\
  & \cso1 \utwoo1' + \fc2(\uoneo0,\utwor) = 0, \eqlabel{u2ep1}
\end{align}
where $\fc{\i\j}(\xf) \equiv \@ \fc{}_\i/\@ \ui\j$ evaluated at
$\u = \us(\xf,0) = \left( \uoneo0(\xf) , \utwor \right)\T$.  The $\O{\eps^1}$
corrections $\cso1$, $\uoneo1$, $\utwoo1$ can be obtained from here in
quadratures, provided that $\fc{12}\not\equiv0$. We will not need the explicit
expressions here, and return to these details when considering the degenerate
case, characterized by $\fc{12}\equiv0$.

The linearization in the comoving frame \eq{linop} is similarly
expanded, $\L = \Lo0 + \eps \Lo1 + \O{\eps^2}$, with
\begin{align*}
  \Lo0 &= \Mx{
    \@_\xf^2 + \fc{11}(\xf) & \fc{12}(\xf) \\
    0 & 0
  }, \\
  \Lo1 &= \Mx{
    \cso1\@_\xf + \fc{111}\uoneo1 + \fc{112}\utwoo1 & \fc{121}\uoneo1 + \fc{122}\utwoo1 \\ 
    0 & \cso1 \@_\xf
  } ,
\end{align*}
where
$\fc{\i\j\k} = \fc{\i\j\k}(\xf) = \@^2\fc\i/\@\ui\j\@\ui\k$
evaluated at $\u=\us(\xf,0) = \left( \uoneo0(\xf) , \utwor \right)\T$,
with the adjoint $\La$ defined by the inner product
$\braket{\efl{}}{\L \efr{}} = \braket{\La \efl{}}{\efr{}}$.  The
eigenfunctions of the operators $\L, \La$ are also expanded for small $\eps$,
$\efr\i = \efro0\i + \eps \efro1\i + \O{\eps^2}$ and
$\efl\j = \eflo0\j + \eps \eflo1\j + \O{\eps^2}$.

The eigenfunctions satisfy $\Lo0\efr\i - \eval\i\efr\i = \O{\eps^1}$ and
$\Lo0\H\efl\j - \conj{\eval\j}\efl\j = \O{\eps^1}$.  In the leading order,
this gives by components
\begin{align*}
  \vio0\i1'' + \fc{11}(\xf) \vio0\i1 &= \evalo0\i \vio0\i1 + \fc{12}(\xf) \vio0\i2, \\
  0 &= \evalo0\i \vio0\i2, \\
  \wio0\i1'' + \fc{11}(\xf) \wio0\i1 &= \conj{\evalo0\i} \wio0\i1, \\
  \fc{12}(\xf) \wio0\i1 &= \conj{\evalo0\i} \wio0\i2 ,
\end{align*}
where $\efr\i=\Mx{\vio0\i1,\vio0\i2}\T$, and
$\efl\i=\Mx{\wio0\i1,\wio0\i2}\T$.

For $\i=1$, $\evalo01=\conj{\evalo01} > 0$, we have
\begin{align*}
  \vio011'' + \fc{11}(\xf) \vio011 &= \evalo01 \vio011, \\
  \vio012 &= 0, \\
  \wio011 &= \vio011, \\
  \wio012 &= \fc{12}(\xf) \wio011 / \evalo01 .
\end{align*}
So we have $\vio011 = \wio011 = \ign$ is the ignition mode of the
critical nucleus solution ($\eps=0$); $\vio012=0$; and generically
$\wio012 \neq 0$, hence the naive assumption holds for $\efr1$ but not
$\efl1$.  Note that we can similarly argue that $\vio0\i2=0$ for all
$\i$ whenever $\evalo0\i\ne0$.

For $\i=2$, $\eval\i=0$, the leading order equations are degenerate
and insufficient for finding the eigenfunctions. The right
eigenfunction is known from symmetry consideration, in particular
\begin{align*}
  \vio021 &= \uoneo0', \\
  \vio022 &= 0, \\
  \vio121 &= \uoneo1', \\
  \vio122 &= \utwoo1', 
\end{align*}
whereas the leading order for the left eigenfunction gives
\begin{align*}
  \wio021 &= 0, \\
  \wio022 &= \wio022(\xf),
\end{align*}
where the last (trivial) equation for $\wio022$ is understood to mean
that any function satisfies the asymptotic eigenproblem at this stage,
so long as $\fc{12}(\xf) \neq 0$.  So again the naive assumption holds
for $\efr2$ but not $\efl2$.

For the sake of comparing the asymptotics with the numerics, we would
like to know the asymptotic order of $\vi12$.  The first-order
correction $\efro11$, using standard perturbation theory, is obtained
as a linear combination of $\efro0\j$ for all $\j\ne1$.  We have seen
that inasmuch as $\evalo0\i\ne0$ for all $\i\ne2$, we have
$\vio0\i2=0$, and besides, $\vio022=0$ from symmetry considerations,
hence we conclude that $\vio112=0$.

Finally, the $\O{\eps^1}$ order for $\efl2$ gives
\begin{align*}
  \wio121'' + \fc{11} \wio121 &= 0, \\
  \fc{12} \wio121 &= \cso1 \wio022'.
\end{align*}
Assuming $\wio022(-\infty) = 0$, we find, up to a normalization
constant,
\begin{align*}
  \wio121(\xf) &= \nuc'(\xf), \\
  \wio022(\xf) &= \frac{1}{\cso1}
                 \int_{-\infty}^{\xf} \fc{12}(\xf') \nuc'(\xf') \, \d\xf'.
\end{align*}

To summarize, the expected scaling of the key ingredients of the
theory in the limit of $\eps\to0$ is:
\begin{align*}
  & \cs = \O{\eps}, \\
  & \uone-\uoner=\O{1}, \qquad \utwo-\utwor=\O{\eps}, \\
  & \vi11 = \O{1}, \qquad \vi12 = \o{\eps}, \\
  & \vi21 = \O{1}, \qquad \vi22 = \O{\eps}, \\
  & \wi11 = \O{1}, \qquad \wi12 = \O{1}, \\
  & \wi21 = \O{\eps}, \qquad \wi22 = \O{1} .  
\end{align*}
The behaviour of these ingredients for the FitzHugh-Nagumo system
obtained numerically is illustrated below in~\fig{fhn-sc}, where we
have used the empirically established scaling $\vi12=\O{\eps^2}$.

Taking into the account the structure of the initial perturbation
given by~\eq{pert}, of practical importance are the ``voltage''
components of the shift selectors, $\cmp\l1$ Using the
definitions~\eq{Phil}, we find
\[ 
  \cmp11(\xf) = \ign'(\xf) + \O{\eps},
\]
\[ 
  \cmp21(\xf) = \nuc'(\xf) + \O{\eps},
\]
and
\[
  \cmp31(\xf) = \A \ign(\xf) + \B \nuc'(\xf) + \O{\eps}, 
\]
where $\A$ and $\B$ are some constants; for reference,
\[
  \A = - \frac{1}{\cso1}
  \iint\limits_{\xf'\le\xf} \fc{12}(\xf') \nuc'(\xf') \utwoo1(\xf) \,\d\xf'\,\d\xf ,
\]
\[
  \B = \intinf \ign(\xf) \left( \nuc(\xf) -\uoner \right)\,\d\xf .
\]
Observe that since $\nuc(\xf)$ and $\ign(\xf)$ are even functions, we
have that $\cmp11(\xf)$ and $\cmp21(\xf)$ are odd in the limit
$\eps\to0$, which guarantees the availability of the choice $\shift=0$
for these selectors, as would be expected.  At the same time, since
$\A$ and $\B$ are typically both nonzero, $\lim_{\eps\to0}\cmp31(\xf)$
is not odd, and the choice $\shift=0$ is not available in this
case. Though the limit $0 \neq \lim_{\eps\to0}\shift$ exists for
$\cmp31(\xf)$.

\subsection{Asymptotic structure for the degenerate case}
\seclabel{asymp-degenerate}

In the case of the Karma model and also for the cubic recovery variant
of the FitzHugh-Nagumo model, the standard asymptotics described above
do not work. More precisely, it fails for any model in which
$\fc{12}(\ui1,\ui2)\equiv0$. To see why, let us consider in more
detail the $\O{\eps^{1}}$ corrections $\cso1$, $\uio21$, and $\uio11$.
From \eq{u2ep1} and the asymptotic boundary condition
$\ui2(\xf\to+\infty) \to 0$ it follows that
\begin{equation}\eqlabel{u2}
  \ui2(\xf) = -\frac{1}{\cso1} \Vci(\xf),
\end{equation}
where
$\Vci(\xf) = \int_{\xf}^{\infty} \, \fc2(\uio10(\xp),\uri2) \, \d
\xp$. This gives the leading order $\O{\eps^1}$ of the slow component
of the critical pulse.  The value of $\cso1$ can be obtained if we
multiply \eq{u1ep1} by $\uio10$ and integrate,
\begin{equation}\eqlabel{c1}
  \cso1 = \left(\frac{-\int_{-\infty}^{+\infty} \,
      \fc{12}(\uio10,\uri2) \Vci \uio10 \, \d
      \xf}{\int_{-\infty}^{+\infty} 
      \left(\uio10\right)^2 
      \d\xf}\right)^{1/2} , 
\end{equation}
by exploiting \eq{u1ep0} and the boundary conditions
$\uio10'(\pm\infty)=\uio11'(\pm\infty)=\fc1(\uio10(\pm\infty))=0$.

In the degenerate case, $\fc{12}(\ui1,\ui2)\equiv0$, and according to
\eq{c1} we have $\cso1=0$, and consequently no answer for ${\uio21}'$
which has $\cso1$ in the denominator in \eq{u2}.  Therefore the
asymptotics are to be determined separately, taking into account the
specific dependence of $\fc1$ on $\ui2$.  We consider the dependence
of the form $\fc1(\ui1,\ui2) = \fd{}(\ui1,\ud)$, where
$\ud(\ui2) = \ui2^\q$ and $\q>0$, so the the problem for the critical
pulse is
\begin{align}
  \ui1'' + \cs \ui1' + \fd{}\left(\ui1,\ui2^\q\right) &= 0, \eqlabel{u1tw} \\
  \cs \ui2' + \gam \fc2(\ui1, \ui2) &= 0,\eqlabel{u2tw}
\end{align}
and postulate $\ui1 = \uio10 + \gam^{\pwu}\uio11$,
$\ui2 = \ui2^{(0)} + \gam^{\pwv}\uio21$, and $\cs = \gam^{\pwc} \cso1$
to leading order.  Substitution into the traveling wave equations
yields the expected equation~\eq{u1ep0} for the critical nucleus
solution, while the second equation relates two terms which must match
to leading order in $\gam$,
\[
  \gam^{\pwc + \pwv}\cso1\uio21 + \gam \fc2(\uio10,\uri2) = 0,
\]
from which we conclude that $\pwc + \pwv = 1$.

Considering the next-to-leading order in $\gam$ from \eq{u1tw}
we have
\begin{align*}
  & \gam^{\pwu}\ui1^{(1)''} + \gam^{\pwc}\cs_1^{(1)}\ui1^{(0)'} + \\
  & \gam^{\pwu} \fd1(\uio10,\uri2)\uio11
    + \gam^{\q \pwv} \fd2(\uio10,\uri2) {\uio21}^\q + \mathrm{h.o.t.} = 0,
\end{align*}
for which the balance to leading order in $\gam$ is achieved for
$\pwu = \pwc = \q \pwv$.  Combined with the previous results for
$\pwv$ and $\pwc$, this gives
\begin{equation}\eqlabel{nonlinearscaling}
  \pwu = \pwc = \frac{\q}{\q+1}, \quad \pwv = \frac{1}{\q+1} .
\end{equation}
Introducing $\eps=\gam^{1/(\q+1)}$, we summarise that the nonlinear
solution scales as
\begin{align*}
  \ui1 &= \uio10 + \eps^\q \uio11 + \mathrm{h.o.t}, \\
  \ui2 &= \uri2 + \eps \uio21 + \mathrm{h.o.t}, \\
  \cs &= \eps^\q\cso1 + \mathrm{h.o.t} .
\end{align*}
While we recover $\pwu=\pwv=\pwc=1/2$ in the classical case $(\q=1)$,
we have more exotic scaling for different values of $\q$.

We now determine the scaling of the solutions of the linearised
problems, focussing on $\q>1$.  The leading terms of the linearization
operator in the comoving frame are
\[
  \L \approx \Mx{1&0\\0&0} \@_\xf^2 + \eps^{\q} \cso1 \Mx{1&0\\0&1}\@_\xf +
  \Mx{\fc{11}& \q \usi2^{\q-1} \fd\ud \\ \eps^{\q+1}\fc{21} & \eps^{\q+1} \fc{22}},
\]
and  of its adjoint
\[
  \La \approx \Mx{1&0\\0&0} \@_\xf^2 - \eps^{\q} \cso1 \Mx{1&0\\0&1}\@_\xf +
  \Mx{\fc{11}& \eps^{\q+1}\fc{21}  \\ \q \usi2^{\q-1} \fd\ud & \eps^{\q+1} \fc{22}},
\]
where the derivatives of the kinetic terms, including
$\fd\ud\equiv\@{\fd{}}/\@\ud$, are understood to be evaluated at the
critical solution.

For the leading order for the first eigenpair we have
\begin{align*}
  & \eval1=\evalo01+\hot, \qquad \evalo01=\O{1},  \\
  & \efr1=\Mx{\vi11 \\ \vi12 }=\efro01+\hot, \\ 
  & \vi11 = \vio011 + \hot, \qquad \vio011=\O{1}, 
\end{align*}
the latter being down to our arbitrary choice of normalization.  In
the leading order, the equation for the first component decouples from
the second equation,
\[
  \vio011'' + \fc{11} \vio011  = \evalo01\vio011 + \O{\eps^\q} + 
  \O{\eps^{\q-1}\vio012} .    
\]
Hence, assuming $\vio012=\O{\eps}$, we expect that
\[
  \vi11=\vio011 + \O{\eps^\q}, \qquad \eval1=\evalo01 + \O{\eps^{\q}} .
\]
The second component is then to be obtained from the second equation,
with the first component and the eigenvalue considered as given:
\[
  \eps^\q\cso1\vi12' 
  + \eps^{\q+1}\fc{21}\vio011 
  + \eps^{\q+1} \fc{22}\vio012 
  = \eval1\vi12 . 
\]
Note that the $\vi12$-dependent terms on the left-hand side are
asymptotically smaller than the right-hand side, hence the balance is
achieved via
\[
  \eval1\vi12 = \eps^{\q+1} \fc{21}\vio011 + \hot , 
\]
so that the leading order contribution of $\vi12$ is
\[
  \vi12 = \O{\eps^{\q+1}} ,
\]
making self-constent our earlier assumption that $\vi12$ does not
exceed $\O{\eps}$.  Curiously, we observe that the leading term in
$\vi12=\O{\eps^{\q+1}}$ is smaller than the first-order correction in
$\vi11$, which is $\O{\eps^\q}$.

The second eigenpair is different in that we know $\eval2=0$ exactly
and $\efr2 = \u'$ due to translational symmetry.  Otherwise we proceed
as before,
\begin{align*}
  & \efr2=\Mx{\vi21 \\ \vi22 }=\efro02+\hot, \\ 
  & \vi21 = \vio021 + \hot, \qquad \vio021=\O{1} ,
\end{align*}
and the equation for the first component gives
\[
  \vio021'' + \fc{11} \vio021  = \O{\eps^\q}
\]
so we expect
\[
  \vi21=\vio021 + \O{\eps^\q},
\]
in accordance with the leading asymptotic expansion of $\ui1$.  Then
the second equation gives
\[
  \eps^\q\cso1\vi22' 
  + \eps^{\q+1}\fc{21}\vio021
  + \eps^{\q+1} \fc{22}\vi22
  = \hot
\]
Comparison of the first and second term here shows that $\vi22$ is of
a higher asymptotic order than $\vi21$.  We therefore can neglect the
third term in comparison with the second, which leads to
\[
  \vi22 = -\frac{\eps}{\cso1} \int \fc{21} \vio021 \, \d\xf + \hot,
\]
such that $\vi22 = \O{\eps}$, echoing the asymptotic expansion of
$\ui2$.

For the first left eigenfunction, we have
\begin{align*}
  & \wi11'' + \fc{11}\wi11 = \eval1\wi11 + \hot\\
  & \q \eps^{\q-1}\uio21^{\q-1} \fd\ud \wi11  
    = \eval1\wi12 + \hot
\end{align*}
Hence we can take $\wio011 = \vio011 = \O{1}$, and then
\[
  \wi12 = \eval1^{-1} \q \eps^{\q-1}\uio21^{\q-1} \fd\ud \wi11
  = \O{\eps^{\q-1}}. 
\]

Finally, for  the second left eigenfunction, we have
\begin{align*}
  & \wi21'' + \fc{11}\wi21 = 0 + \hot\\
  & -\eps^\q\cso1\wi22' 
    + \q \eps^{\q-1}\uio21^{\q-1} \fd\ud \wi21  = 0 + \hot,
\end{align*}
and therefore
\begin{align*}
  & \wio021 = \vio021 = \O{1} \\
  & \wi22 = \eps^{-1} 
    ( \q/\cso1 ) \int \uio21^{\q-1} \fd\ud \wi21 \,\d\xf + \hot
    = \O{\eps^{-1}} .
\end{align*}

To summarise, the eigenfunction components scale as
\begin{align*}
  & \vi11 = \O{1}, && \vi12 = \O{\eps^{\q+1}}, \\
  & \vi21 = \O{1}, && \vi22 = \O{\eps}, \\
  & \wi11 = \O{1}, && \wi12 = \O{\eps^{\q-1}},  \\
  & \wi21 = \O{1}, && \wi22 = \O{\eps^{-1}} . 
\end{align*}
Note that the generic case asymptotics are recovered by setting $\q=1$
and correspondingly $\eps=\gam^{1/2}$, including the scaling of
$\vi12$ which in the generic case was not established conclusively.

%%%%%%%%%%%%%%%%%%%%%%%%%%%%%%%%%%%%%%%%%%%%%%%%%%%%%%%%%%%%%%%%%%%%%%%%%%%%%%%

\section{Methods}

Throughout the remainder of this paper we shall deal with two-variable
systems of partial differential equations of the form given in
\eq{rde}, distinguished by the details of the functional form of
$\f$.  The essential ingredients of the linear theory for predicting
the critical excitation strength-extent relationship remain the same
across these models, however, and our methods for computing these are
likewise similar.  We begin by writing the system of partial
differential equations in the frame moving with speed $\c$ yields a
system of three ordinary differential equations,
\begin{eqsplit}{ode}
  \ui1' &= \ui3, \\
  \ui2' &= \c^{-1} \gam \fc2(\ui1,\ui2), \\
  \ui3' &=  (-\c \ui3 + \fc1(\ui1,\ui2,\ui1') + \Iex), 
\end{eqsplit}
whose unique equilibrium is given by $[\ur,0]$.  Continuing the rest
state for increasing current forcing $\Iex$ connects to a Hopf
bifurcation, from which a family of periodic orbits emanate.
Continuing this family of periodic orbits to large periods with
$\c \neq 0$ followed by decreasing current forcing yields an unforced
periodic orbit of the autonomous system, \eq{ode}, equivalently
a traveling wave solution of \eq{rde} with periodic boundary
conditions on a domain $\x\in [0,\Length)$.  To compute
\emph{asymptotic} traveling wave solutions of \eq{rde} we
continue the unforced periodic orbit in the $(\Length,\c)$-plane.  The
asymptotic solutions correspond to large domain sizes, for which $\c$
becomes constant:
$\lim_{\Length\to +\infty} \c'(\Length) = 0$.  In this
asymptotic regime, $\lim_{\Length\to +\infty} \c(\Length)$ is
multi-valued, and the lowest of the speeds,
$\c=\cs$,
designates the critical
solution, $(\us,\cs)$.  In the large-$\Length$ limit the periodic
solution approximates the homoclinic originating from the rest state,
and as a practical matter this limit is numerically inaccessible,
particularly for small $\gam$.  In this context, we will approximate
the homoclinic solution with projection boundary
conditions~\cite{Beyn-1990}, which permits aperiodic solutions
by enforcing orthogonality to the stable/unstable eigenspaces of the
rest state at the boundaries of the domain.

The periodic critical solution is computed on an adaptive collocation
grid using \Auto~\cite{Doedel-Kernevez-1986}, at large $\gam$ and interpolated onto a
Chebyshev grid of size $\Modes\times\DA$ representing $\Modes$
Chebyshev modes and a dealiasing factor $\DA \geq 1$.  A nonlinear
boundary value problem is constructed which corresponds to equations
\eq{ode} and projection boundary conditions.  The projection
boundary conditions require the eigenvectors of the Jacobian evaluated
at the rest state.  The Jacobian of the ODE system is given by
\begin{equation}\eqlabel{odejac}
  \Jac = \Mx{
    0 & 0 & 1 \\ 
    -\gam \fc{2,1}/\cs & -\gam \fc{2,2}/\cs & 0 \\ 
    -\fc{1,1} & -\fc{1,2} & -\cs
  }
\end{equation}
with unstable and stable subspaces spanned by the right eigenvectors
satisfying $\Mx{\Eu,\Es}\diag{\Lamu,\Lams} = \Jac\,\Mx{\Eu,\Es}$.  We
require that the unstable pulse travels to the right ($\cs>0$) and thus
that the perturbation from the rest state on the right hand side be
orthogonal to the stable subspace (guaranteeing excitation dynamics),
while the perturbation from the rest state on the left hand side be
orthogonal to the unstable subspace (guaranteeing a relaxation to the
rest state).  The projectors of $\Eu$ and $\Es$ are the corresponding
left eigenvectors of $\Jac$, so that
$\Mx{\Pu,\Ps}^\top = \Mx{\Eu,\Es}^{-1}$, and the boundary conditions
are
\begin{align*}
  \Pu^\top \Mx{\us(0)-\ur,\ui1'(0)} &= 0, \\
  \Ps^\top \Mx{\us(\Length)-\ur,\uone'(\Length)} &= \0,
\end{align*}
where $\Pu$ gives one condition at the left boundary ($\x=0$) and
$\Ps$ gives two conditions at the right boundary ($\x=\Length$).  The
boundary value problem is solved using the Newton solver in the
open-source \Dedalus\ framework~\cite{Burns-etal-2019} until the update is
smaller than the tolerance of $\etol = 5\E{-13}$ in $\Linf$-norm.

The linearization utilizes the boundary value problem solution within
the forcing terms, and is likewise discretized using $\Modes\times\DA$
grid points $\Modes$ Chebyshev modes and a dealiasing factor of
$\DA \geq 1$.  The projection boundary conditions for the
linearization follow the same logic as presented for the nonlinear
boundary value problem.  The eigenproblem is solved by calling the
sparse eigensolver package \Arpack\ through \code{scipy.linalg.eigs},
with up to $4096$ iterations retaining $64$ Lanczos basis vectors to
resolve the leading eigenmodes (significantly fewer than $\Modes$).
Note that the adjoint linearization equations and adjoint boundary
conditions are distinct from the forward linearization equations and
boundary conditions; in particular, while the forward linearization
maintains one boundary condition at the left boundary and two
conditions at the right boundary, the adjoint linearization problem
applies two conditions at the left boundary and one condition at the
right boundary.  The Jacobian for the forward eigenproblem is the same
as the nonlinear problem Jacobian, \eq{odejac}, and thus uses
the same boundary conditions applied to the first-order form of
$\efr\i$.  The Jacobian for the adjoint eigenproblem is given by
\[
  \Jac\H = \Mx{
    0 & 0 & 1 \\ 
    \fc{1,2}/\cs & \gam \fc{2,2}/\cs & 0 \\ 
    -\fc{1,1} & -\gam \fc{2,1} & \cs
  },
\]
such that
$\Jac\H \Mx{\Eu\H,\Es\H} = \Mx{\Eu\H,\Es\H} \diag{\conj{\Lamu},
  \conj{\Lams}}$, and $\Mx{\Pu\H,\Ps\H}\T = \Mx{\Eu\H,\Es\H}^{-1}$,
and the corresponding boundary conditions are
\begin{align*}
  {\Pu\H}\T \Mx{\efl{}(0),\w_1'(0)} &= \0, \\
  {\Ps\H}\T \Mx{\efl{}(\Length),\w_1'(\Length)} &= 0 .
\end{align*}

Since the forward and adjoint linearized eigenproblems are formulated
independently, the eigenvalues ($\conj{\eval\j}$) resulting from the
calculation of the left eigenfunctions ($\efl\j$) and the eigenvalues
($\eval\i$) resulting from the calculation of the right eigenfunctions
($\efr\i$) are compared and matched pairwise, and the left and right
sets of eigenfunctions are used to verify the biorthogonality
conditions, $(\eval\j - \eval\i)\braket{ \efl\j }{ \efr\i } = 0$.

The prediction of the critical curve is conveniently done using the
Fourier transform, so the solutions are sampled on a uniform grid of
sufficiently large size $\Ngrid$, for the nonlinear as well as the
linear problems.  It begins by forming the shift selector $\Cmp\l$,
explicitly computing the inner products and appropriate sums using the
normalized eigenfunctions, such that
$\braket{ \efl\j }{ \efr\i } = \kron{\i\j}$.  To determine the shift
value for a prescribed perturbation shape, $\Xb(\x)$, two
cross-correlation integrals are computed.  First, the perturbation is
cross-correlated with $\Cmp\l(\x)$ using the product of the Fourier
transforms,
\[
  \Q\l(\shift) = \intinf \Cmp\l(\xf)\T \Xb(\xf+\shift) \,\d\xf
  = \sqrt{2\pi} \Fti{ \Cmpf\l\H(\qq) \Xbf(\qq) }
\]
where $\Cmpf\l(\xf)(\qq)=\Ft{\Cmp\l(\xf)}$, $\Xbf(\qq)=\Ft{\Xb(\xf)}$,
and the Fourier transform and its inverse are defined as
\begin{align*}
  & \fff(\qq) = \Ft{\ff(\xf)} = \frac{1}{\sqrt{2\pi}} \intinf \ff(\xf)\,\ee{-\ii\qq\xf}\,\d\xf , \\
  & \ff(\xf) = \Fti{\fff(\qq)} = \frac{1}{\sqrt{2\pi}} \intinf \fff(\qq)\,\ee{\ii\qq\xf}\,\d\qq .
\end{align*}
The roots of $\Q\l$ are computed by checking for sequential
differences in the sign of the elements of $\Q\l$, and refined using a
Newton method applied to a locally adapted spline interpolant of
$\Q\l$.  The linear prediction for the critical strength as a function
of the shift~\eq{Us} is also computed using cross-correlation via Fourier
transform,
\[
  \Us(\shift) = \braket{ \efl1(\xf) }{ \us(\xf)-\ur } / \Den(\shift),
\]
where
\[
  \Den(\shift) = \sqrt{2\pi}  \Fti{ \eflf1\H(\qq)\Xbf(\qq) } , 
\]
and selecting the global minimum of $|\Us(\shift)| > 0$ for refinement by
applying a scalar Newton method to a function which computes
$\Q\l(\shift) = 0$ to sub-grid accuracy by a cubic interpolant.  The
root of $\Q\l(\shift)$ determines the position of the perturbation at
time $\t=0$, $\Xb(\xf+\shift) = \Xb(\x)$, where $\xf=\x-\cs\t-\shift$.

The linear theory prediction is compared to direct numerical
simulations (DNS), using initial conditions corresponding to the
perturbation shape, $\u(0,\x) = \ur + \Us\Xb(\x)$.  The problem is
solved on the domain $\x \in [0, \length]$ where the domain size is no
larger than half the asymptotic domain size $\length \leq \Length/2$,
and over a time interval $ \t \in [0, \Time]$ where $\Time$ is
comparable to $\cf/\length$ where $\cf$ is the speed of the fastest
isolated asymptotic wave.  The problem is discretized in space using a
Chebyshev spectral method using $\Modes$ modes, and time-stepped using
a third-order implicit-explicit method (RK443~\cite{Ascher-etal-1997}), as
implemented in the numerical package \Dedalus~\cite{Burns-etal-2019}.  The
boundary conditions are chosen as
$\@_\x\ui1(\t,0) = \@_\x\ui1(\t,\length)=0$, and enforced using the
Chebyshev-tau method~\cite{Ortiz-1969}.

For an initial condition parameterized by $\Us$, a wave may ignite and
recruit the entirety of the tissue ($\Us > \Usc$), or it may
immediately decay and approach the rest state ($\Us < \Usc$).  To
numerically distinguish between these events it is necessary to track
the state as it evolves over time, and in particular for the tracking
to unambiguously characterize these outcomes.  To this end we define a
distance function,
\begin{multline*}
  \dist(\t)
  \equiv \intinf \abs{\eb\T (\u(\t,\x) - \ur)} \,\d\x \\
  - \intinf \abs{\eb\T (\u(0,\x) - \ur) } \,\d\x ,
\end{multline*}
which compares the amplitude of the initial condition in the voltage
channel to the amplitude of the state in the voltage channel at all
later times.  Considering only the final value $\dist(\Time)$ gives an
effective scalar function, $\ff: \Real \to \Real$, and for
sufficiently smooth flows $\ff$ is continuous.  For initial conditions
set to the critical wave solution, $\u(0,\x) = \us(\x)$, it is clear
that $\dist(\t) = 0$, as it is indeed for all initial conditions which
are equilibria of the underlying partial differential
equations~\eq{rde}.  For generic initial conditions
parameterized by $\Us$, and considering an interval
$\Us \in [\Usl, \Usu]$ where $\ff(\Usl) < 0$ and $\ff(\Usu) > 0$, we
can compute the roots of $\ff$ using a iterative bisection procedure.
Additionally parameterizing the perturbation to the rest state by the
width, $\xs$, then for each sampled width $\xs$ the range of $\Us$ is
determined, and the bisecting procedure is applied to determine the
pair $(\xs, \Usc)$ which defines the critical perturbation.

\section{Results}

Throughout this section we compare DNS results to the predictions
computed using the linear theory, and pay particular attention to the
variation in the critical amplitude predictions for different choices
of shift-selecting heuristics, as well as the adequacy of the
predictions as we approach the singular limit of these slow-fast
systems parametrically.  We also compare the critical amplitude to the
unstable root of the kinetics, $\fc1(\ui1,\utwor) = 0$, which is the
asymptotic threshold for the double limit $\gam\to 0$, $\xs\to\infty$
for the class of perturbation shapes used in this work.  We perform
these comparisons for a small but representative set of nonlinear
excitation models with two variables, with temporal and structural
differences but falling under the slow-fast paradigm.

\subsection{FitzHugh-Nagumo}

FitzHugh-Nagumo is a prototypical model of excitation following the
asymptotic reduction of the Hodgkin-Huxley model equations for the
giant squid axon~\cite{FitzHugh-1961}.  The FitzHugh-Nagumo kinetics
for $\f$ are given below,
\begin{align}\eqlabel{fhn}
  \fc1 &= \ui1(1-\ui1)(\ui1-\fb) - \ui2,\\
  \fc2 &= \fa \ui1 - \ui2, \nonumber
\end{align}
with fixed $\fa = 0.37$, $\fb= 0.131655$.  The speed ratio varied in
the interval $10^{-10} \leq \gam \leq 10^{-2}$.  The precise choice of
$\fb$ is such that it equals to its critical value
$\fbc=\gamc +2\sqrt{\fa\gamc}$, which corresponds to the transition
between stable node and stable focus in the local kinetics, for
$\gam=\gamc=10^{-2}$.

The first step is to compute the unstable wave solution, $\us(\x)$,
and then its linear eigenspectrum.  We repeat these calculations for
varied $\gamma$, from critical $\gamma=\gamma_c$ down to very small
value to test the singular limit.  Note that the singular limit of
traveling waves in the classical FitzHugh-Nagumo system is
well-studied~\cite{%
  Guckenheimer-Kuehn-2009,%
  Flores-1991,%
  Tyson-Keener-1988,%
  Casten-etal-1975%
}.

\sglfigure{\includegraphics{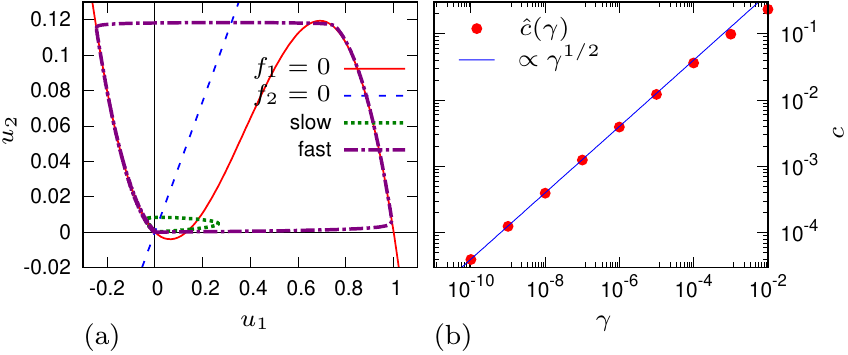}}{%
  Propagating waves in the FHN model~\eq{fhn}. (a) Phase plane of the
  point system with the null-clines and trajectories corresponding to
  the stable and unstable pulses, for $\gam=10^{-3}$. (b) Speed of the
  critical (slow) pulse as function of $\gam$, together with the
  theoretical asymptotic.%
}{fhn-nlin}

The scaling results of the previous section apply directly to this
model, and thus serve as a method of verifying the numerical results
under variations in the modal expansion length $\Modes$, domain length
$\Length$, and the application of projection boundary conditions,
across several decades in $\gam$.  In particular, we find the expected
asymptotic scaling of the wave components and pulse speed with $\gam$
for $\Length\geq 50$ and $\Modes\geq 32$, with exponentially small
corrections to the leading eigenvalues for $\Modes\geq 256$ when
$\Length \geq 100$.  For sufficiently small $\gam$, the solution of
the marginal eigenvalue $\eval2=0$ is eventually corrupted by the
proximity of the essential spectrum of the wave to this eigenvalue,
and the eigenfunctions $\efr2$ and $\efl2$ may not be computed using
these methods; while $\Modes$ and $\Length$ dependent, we have found
this to occur for $\gam < 10^{-10}$, placing a limit on the
reliability of the numerical approach taken in this work.
\Fig{fhn-nlin} demonstrates the correct scaling of the pulse speed
$\cs$ with $\gam$, with good agreement with the $\gam^{1/2}$ behavior
for small $\gam$.

\dblfigure{\includegraphics{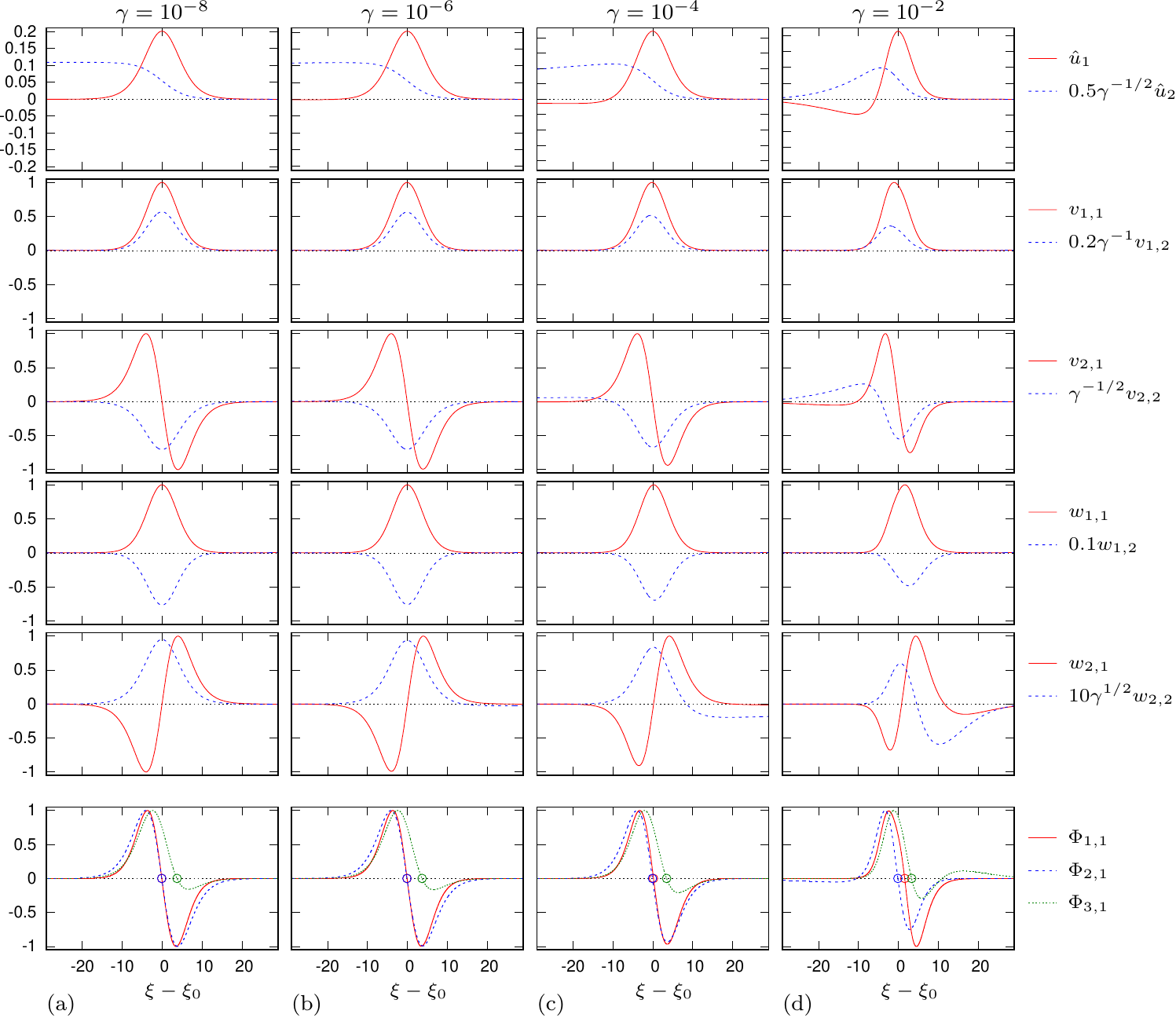}}{%
  Scaled ingredients of the linearized theory for the FHN
  model~\eq{fhn} with varying $\gam$ as specified above the columns.
  The coordinate $\xf=\xo$ corresponds to the position of the maximum
  of $\usi1$.  The open circles the panels in the bottom row indicate
  position of the zeros of the corresponding shift selectors. %
}{fhn-sc}

\Fig{fhn-sc} summarizes the scaling of the components of the linear
theory across six decades of $\gam$, additionally showing the fast
component of the constructed $\Cmp{\l}$.  Note that since the
perturbation is along the fast component, only the fast components of
the shift selector, $\cmp\l1$, are relevant.  Despite the correct
observed scaling of the critical pulse $\us$ and leading
eigenfunctions $\efr1$, $\efr2$, $\efl1$, $\efl2$, the construction of
$\Cmp{\l}$ indicates a subtlety in the selection of an optimal frame
for the prediction of critical perturbations.  While $\Cmp{1}$ and
$\Cmp{2}$ converge to antisymmetric functions centered about the
maximum of $\usi1$, $\Cmp{3}$ instead converges to an asymmetric
function whose central root remains offset from $\xf=\xf_0$ in the
limit $\gam\to 0$.  Considering the components of $\Cmp{3}$ as defined
in \eq{Phil}, we recognize that the contribution proportional to
$\efl1$ does not monotonically vanish as $\gam \to 0$, as originally
expected in the formulation of the linear theory
\cite{Bezekci-Biktashev-2017} which assumed identity of the $\gam\to0$
limit with the $\gam=0$ ``critical pulse'' case.

\dblfigure{\includegraphics{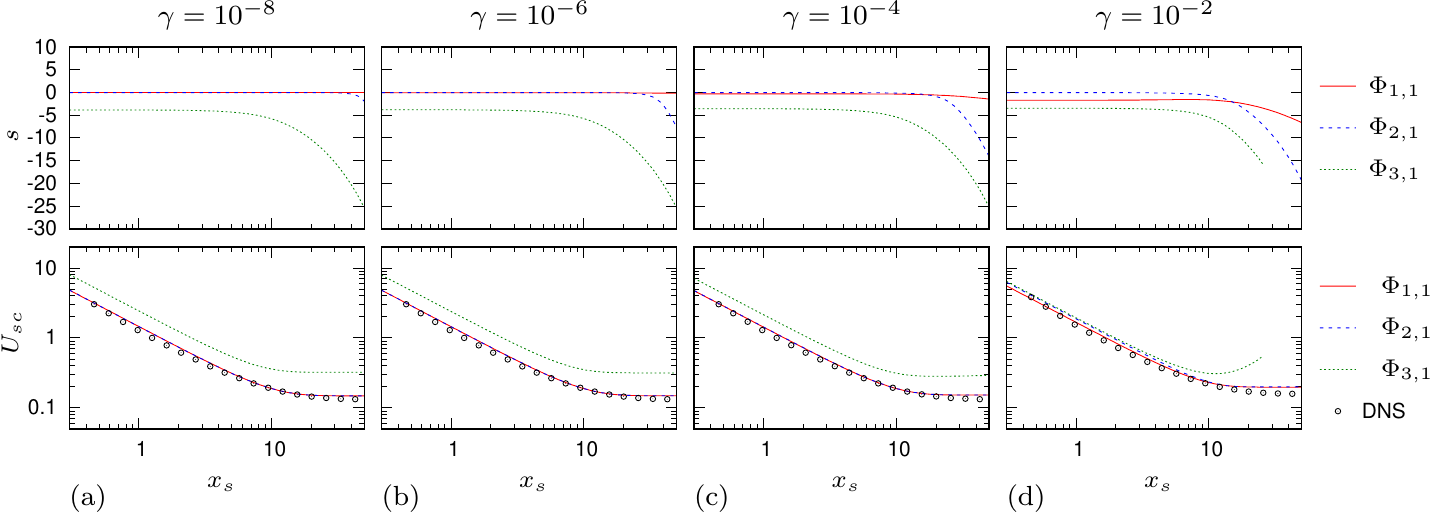}}{%
  Shifts $\shift$ predicted by the three different shift selectors
  $\Cmp\l$ (upper row), and the corresponding predicted critical
  curves against the direct numerical simulations (lower row), for the
  FitzHugh-Nagumo model~\eq{fhn}. %
}{fhn-ccs}

The deviation in the central root of $\Cmp{3}$ has implications for
the selection of optimal reference frames and the critical
perturbation amplitudes associated with them.  \Fig{fhn-ccs} shows the
critical strengths over three decades of the parameterized width
$\xs$, both predicted using $\Cmp{\l}$ and computed using direct
numerical simulation.  The figure also shows the selection of the
optimal frame shift $\shift$ for each of the predictions, in
particular the systematic offset of $\shift_3$ as $\xs\to 0$, which
coincides with the central root of $\Cmp{3}$.  This systematic offset
in the selection of the optimal frame becomes more severe as
$\gam \to 0$, indicating that the prediction due to $\Cmp{3}$ does not
converge (this is not evident from the figure as presented but
confirmed by careful observation of raw data).  We conclude that as
far as $\Cmp3$ is concerned, the slow-fast system with $\gam \to 0$
and the single-variable fast system dynamics of $\ui1$ in isolation
are qualitatively different, i.e., this is a singular limit.  Further,
it indicates that predictions made at large $\gam$ are in fact more
accurate than those made for smaller $\gam$, which is reflected in the
figure.  Notably, for all observed values of $\gam$, $\Cmp{3}$ fails
to recover the large ($\xs\to\infty$) perturbation limit.  A
superficial interpretation of this small paradox is that in the
$\gam\to0$ limit, all the events that decide the fate of a particular
perturbation happen at the time scale $\O{1}$, so the slow variable
remains almost at its resting value, whereas the heuristic behind
$\Cmp3$ is based on the matching of the initial condition against the
critical pulse in full, i.e. taking into account both components. In
the critical pulse, the slow variable is different from the resting
value; although this difference is small, the sensitivity of the pulse
position to the perturbation in the slow component, measured by
$\wi22$, is on the contrary large in this limit, hence the overall
contribution of the slow component does not vanish in the limit
$\gam\to0$.

A small, but nonetheless important note is that the predictive power
of the linear theory is sensitive to the monotonicity and the root
structure of $\Cmp{\l}$ for some perturbation widths, to the extent
that the dependence of predicted $\shift$ and $\Us$ on $\xs$ is
discontinuous.  Such discontinuity occurs, for example, when a branch
of $\Q\l(\shift)=0$, while continuing in $\xs$, terminates in a fold.
In the vicinity of the fold value of $\xs$, the solution
$(\shift,\Us)$ on one side of the fold may differ significantly from
the solution determined on the other side of the fold.
    
This appears in the linear theory prediction utilizing $\Cmp{3}$,
i.e. \fig{fhn-ccs}, for $\gam=10^{-2}$.  For $\xs \gtrsim 25$ the
linear theory makes no reasonable predictions for $\Us$, and for
$5 \lesssim \xs \lesssim 25$ the prediction for $\Us$ diverges from
the asymptotic value correctly predicted by $\Cmp{1}$ and $\Cmp{2}$.
The mechanical reason for this deviation is that the correlation
integral defining $\Q\l$ corresponds to a box filter or smoothing
operation, and for sufficiently large $\xs$ this smoothing is
destructive.  This smoothing destroys the delicate root structure seen
in $\Cmp{3}$ for $\gam=10^{-2}$; so that while for $\xs \to 0$ the
correlation integral forms a translation operation (preserving the
number of roots), when the box filter width $\xs$ is comparable to the
distance between successive roots of a function, the result of the
correlation integral may have fewer roots, and specifically, it may
only have roots which produce anomalously large predictions for $\Us$.

\subsection{Mitchell-Schaeffer}

The Mitchell-Schaeffer~\cite{Mitchell-Schaeffer-2003} model is a popular
semi-conceptual model of cardiac cells, combining the simplicity of
only two components with the relatively realistic description of
action potential shape and restitution properties. Historically, it
has been derived via an asymptotic reduction (adiabatic elimination of
the fastest processes) of the more detailed Fenton-Karma model of
atrial excitation, but still incorporates multiple decay timescales
for the system.  We re-write the kinetics of the model in the form
\begin{align}\eqlabel{mitsch}
  \fc1 &= (1-\ui1)\ui1^2\ui2 - \ui1\mti/\mtu, \\
  \fc2 &= \left((1-\mth(\ui1))(1-\ui2)(\mtc/\mto) - \mth(\ui1) \ui2  \right), \nonumber
\end{align}
where
$\mth(\ui1) = \Heav_\mk(\ui1-\mug) =
\left\{1+\tanh\left[\mk(\ui1-\mug)\right]\right\}/2$ is a smoothed
Heaviside distribution centered at $\ui1=\mug$ with width $\mk^{-1}$,
and the timescales ratio in terms of the original parameters is
$\gam = (\mti/\mtc)$ in this rescaling.  The standard parameter values
are $\mti=0.3\,\ms$, $\mto=120\,\ms$, $\mtu=6\,\ms$, $\mtc=150\,\ms$,
$\mug = 0.03$ and $\mk=100$.  In the following examples, the parameter
ratios $\mti/\mtu=0.05$ and $\mtc/\mto=1.25$ are kept at standard
values, while $\gam$ is treated as a free parameter. That is, for the
purpose of the asymptotic theory, all of $\mti/\mtu$, $\mug$ and
$\mk^{-1}$ are treated as finite even though they are ``small'' in
layman's terms.  The time $\t$ is dimensionless as presented, likewise
we absorb the original diffusion coefficient in the non-dimensional
spatial scale, $\x$.

\sglfigure{\includegraphics{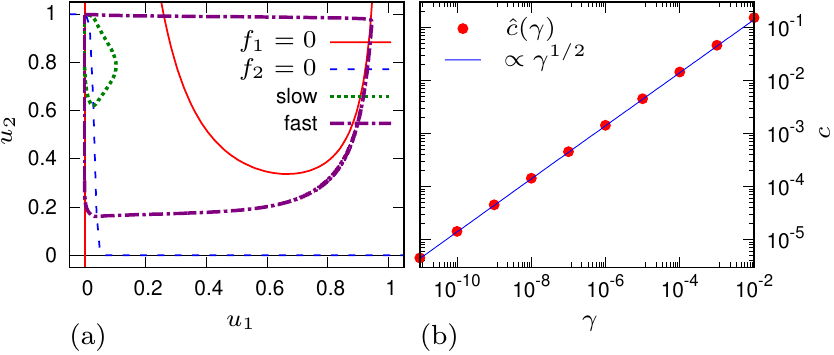}}{%
  Propagating waves in the Mitchell-Schaeffer model~\eq{mitsch}.  (a)
  Phase plane of the point system with the null-clines and
  trajectories corresponding to the stable and unstable pulses, for
  $\gam=10^{-3}$. (b) Speed of the critical (slow) pulse as function
  of $\gam$, together with the theoretical asymptotic. %
}{ms-nlin}

\dblfigure{\includegraphics{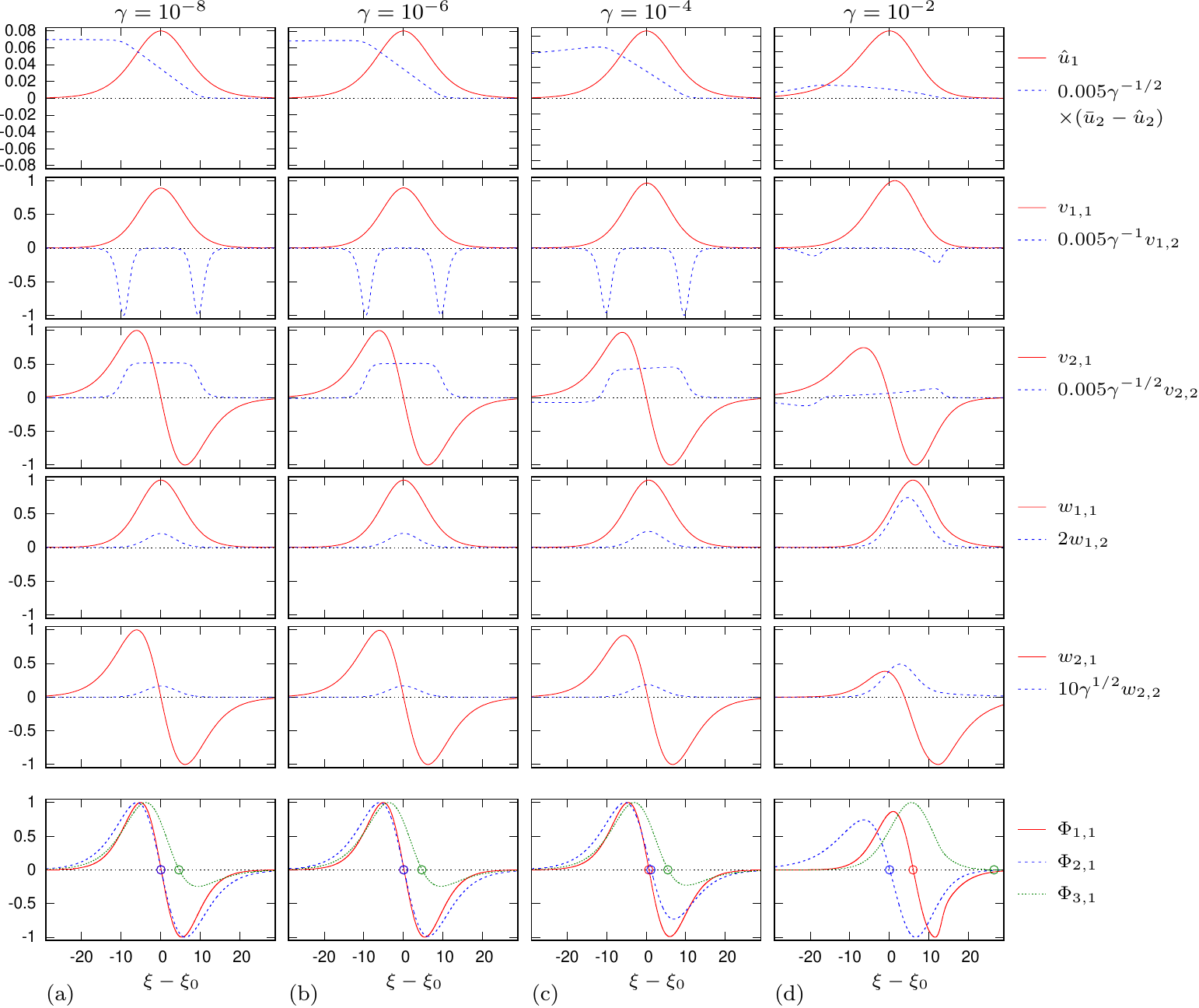}}{%
  Scaled ingredients of the linearized theory for the
  Mitchell-Schaeffer model with varying $\gam$ as specified above the
  columns.  The coordinate $\xf=\xo$ corresponds to the position of
  the maximum of $\usi1$.  The thin vertical lines on the panels in
  the bottom row indicate position of the zeros of the corresponding
  shift selectors. %
}{ms-sc}

The scaling of the critical pulse and associated leading
eigenfunctions in the Mitchell-Shaeffer model follow the expected
scaling, see \fig{ms-nlin} and \fig{ms-sc}, similar to the
FitzHugh-Nagumo model, while reproducing more realistic action
potentials and gate switching dynamics.  However, while the
localization of the critical pulse in the fast variable ($\ui1$) is
not dissimilar to the critical pulse in FitzHugh-Nagumo, the relative
scale of the components of the leading eigenfunctions are reversed.
Namely, we note that in relative terms, the second components of
$\efr1$ and $\efr2$ are two orders of magnitude larger than would be
expected based on their asymptotic order in $\gam$ alone. This is
related to the above mentioned non-asymptotic small parameters in the
model, specifically, sharp switching of the $\fc2$ across $\ui1=\mug$.

The shape of $\Cmp1$ and $\Cmp2$ are standard -- both are
approximately antisymmetric about the peak of the wave, see the bottom
row in \fig{ms-sc}.  However, the contribution in the term
proportional to $\efl1$ in $\Cmp3$ is again significant, so we should
expect a similar offset root and likewise inaccurate predictions for
this shift selector.

\dblfigure{\includegraphics{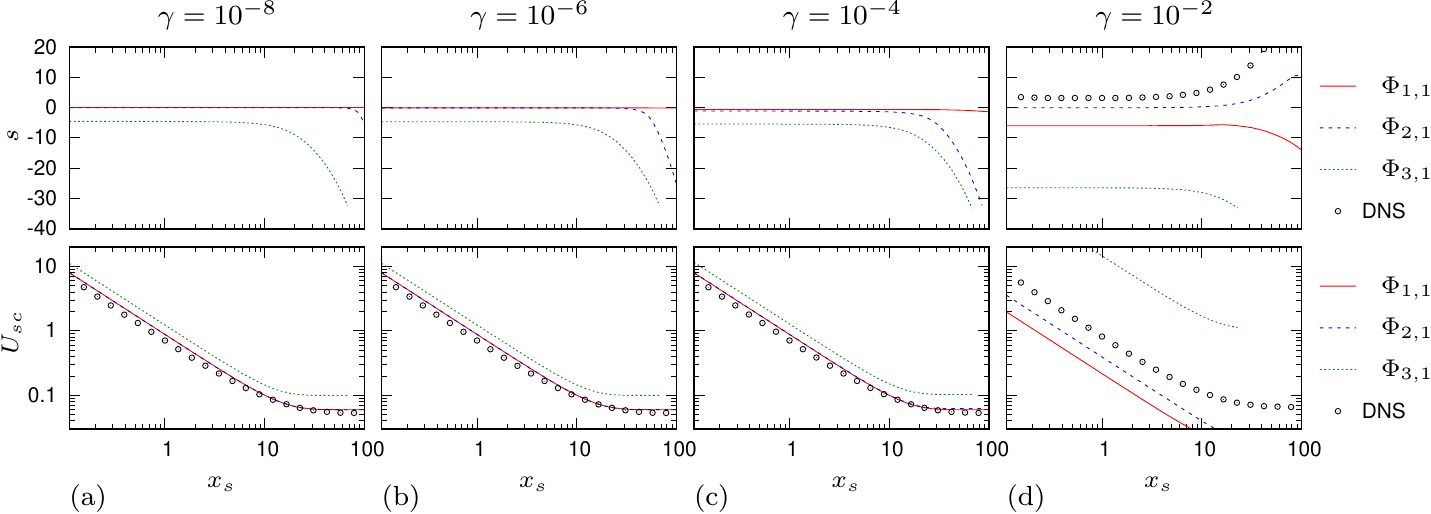}}{%
  Shifts $\shift$ predicted by the three different shift selectors
  $\Cmp\l$ (upper row), and the corresponding predicted critical
  curves against the direct numerical simulations (lower row),
  Mitchell-Schaeffer. %
}{msccs}

The predicted critical excitation amplitudes $\Usc$ track the DNS
results over three orders of magnitude for the extent of the
perturbation, for the value computed using shift selectors $\Cmp1$ and
$\Cmp2$.  As expected, the predictions using the shifts determined by
$\Cmp3$ are much less accurate: they are systematically larger than
the DNS results by nearly an order of magnitude. Note that this error
in $\Usc$ is caused by a deviation of $\shift$ by only a fraction of
the critical pulse width off the predictions of the other two shift
selectors.

Comparison of the predicted results for large $\gam$ indicates that it
is the proximity of the central root of $\Cmp\l$ to the position of
the peak of the unstable pulse which determines the effectiveness of
the prediction, c.f., \fig{ms-sc}.  However, careful examination of
\fig{msccs} indicates that between $\gam=10^{-4}$ and $\gam=10^{-2}$
the primacy of the different shift selectors changes; that is, at
$\gam=10^{-4}$, we see that $\Cmp{2}$ is the most accurate shift
selector, while at $\gam=10^{-2}$ it is $\Cmp{1}$ though no shift
selector generates satisfactorily accurate predictions.

Taking note of \fig{msccs} for $\gam=10^{-2}$, we observe that
$\Cmp{3}$ predicts critical excitation strengths which fail for
$\xs > 11$, i.e., for some sufficiently wide perturbations the branch
of $\Q3(\shift)=0$ ends in a fold, just as with the FHN model results.
This mechanism thus appears to be a generic feature of qualitatively
similar inputs to the infrastructure of the linear theory.  It remains
to determine the conditions under which $\cmp31$ changes shape
sufficiently that the convolution with the perturbation in the
formation of $\Q3$ destroys the optimal root.  In principle, we may
consider a toy model of the dependence $\cmp31(\shift;\gam)$, in the
form
$\toyphi(\shift;\toygam) = \toygam \wi11(\shift) + \sqrt{1-\toygam^2}
\wi21(\shift)$, whereby $\toyphi(\shift;1)$ has no roots, while
$\toyphi(\shift;0)$ has a single root which persists under convolution
with the perturbation $\Xbf$ with a fixed, large, width $\xs$.  At
some intermediate value of $\toygam = \toygam_c$, the number of roots
in the convolution changes, which may be distinct from the value of
$\toygam = \toygam^c$ at which the number of roots within a central
region of the domain of $\toyphi(\,\cdot\,;\toygam)$ changes.  The
disparity between $\toygam_c$ and $\toygam^c$ suggests an analogous
liminal region in which the time-scale separation $\gam$ yields
$\cmp31$ with an appropriate root, but no corresponding root in $\Q3$.

Indeed, \fig{msccs} suggests that for a given $\xs$, we may select a
frame which precisely reflects the DNS results, that is, the actual
position of the critical nucleus observed as a long transient when the
perturbation magnitude is at its closest to the critical value. We say
that this makes a \emph{postdictive} optimal frame through the
determination of $\shift\DNS$, effectively.  The relevant value of
$\gam=10^{-2}$ presents a situation in which all three predictive
curves are distinct and each badly represents the DNS results.
Iterating through perturbation extents we find that
$\shift\DNS > \shift_2 > \shift_1 > \shift_3$ over three decades of
$\xs$.  In the limit of $\xs\to 0$, $\lim\shift\DNS \approx +3.47$,
while $\lim\shift_1\approx -5.94$, $\lim\shift_2\approx0.00$, and
$\lim\shift_3\approx-26.4$.  This exercise informs about the
neighborhood of the predictive measures, however extending this
observation to a general principle, i.e. generating a $\Cmp{}\DNS$, is
far from obvious.

\subsection{Modified ``cubic recovery'' FitzHugh-Nagumo}

\sglfigure{\includegraphics{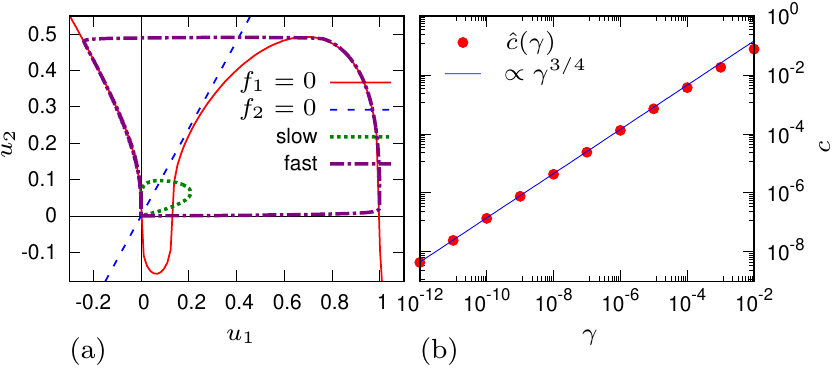}}{%
  Propagating waves in the FHNCR model~\eq{fhnc}. (a) Phase plane of
  the point system with the null-clines and trajectories corresponding
  to the stable front and the unstable pulse, for $\gam=10^{-3}$. (b)
  Speed of the critical (slow) pulse as function of $\gam$, together
  with the theoretical asymptotic. %
}{fnc-nlin}

Here we consider a modification of the FitzHugh-Nagumo model~\eq{fhn}
which is motivated by the nonlinear dependence of $\fc1$ on $\ui2$ in
the Karma model considered in the next subsection.  Our modified FHN
model is
\begin{align}\eqlabel{fhnc}
  \fc1 &= \ui1(1-\ui1)(\ui1-\fb) - \ui2^\q,\\
  \fc2 &= \fa \ui1 - \ui2.  \nonumber
\end{align}
Obviously, \eq{fhn} corresponds to the case $\q=1$.  In this
subsection, we shall look at $\q=3$ instead.  We refer to this model
as ``FHN with cubic recovery'' or FHNCR for short.  We keep
$\fb=0.131655$ as in \eq{fhn}, but take $\fa=1.2$ in order to keep the
kinetics excitable, i.e. have only one equilibrium, see
\fig{fnc-nlin}(a).

\dblfigure{\includegraphics{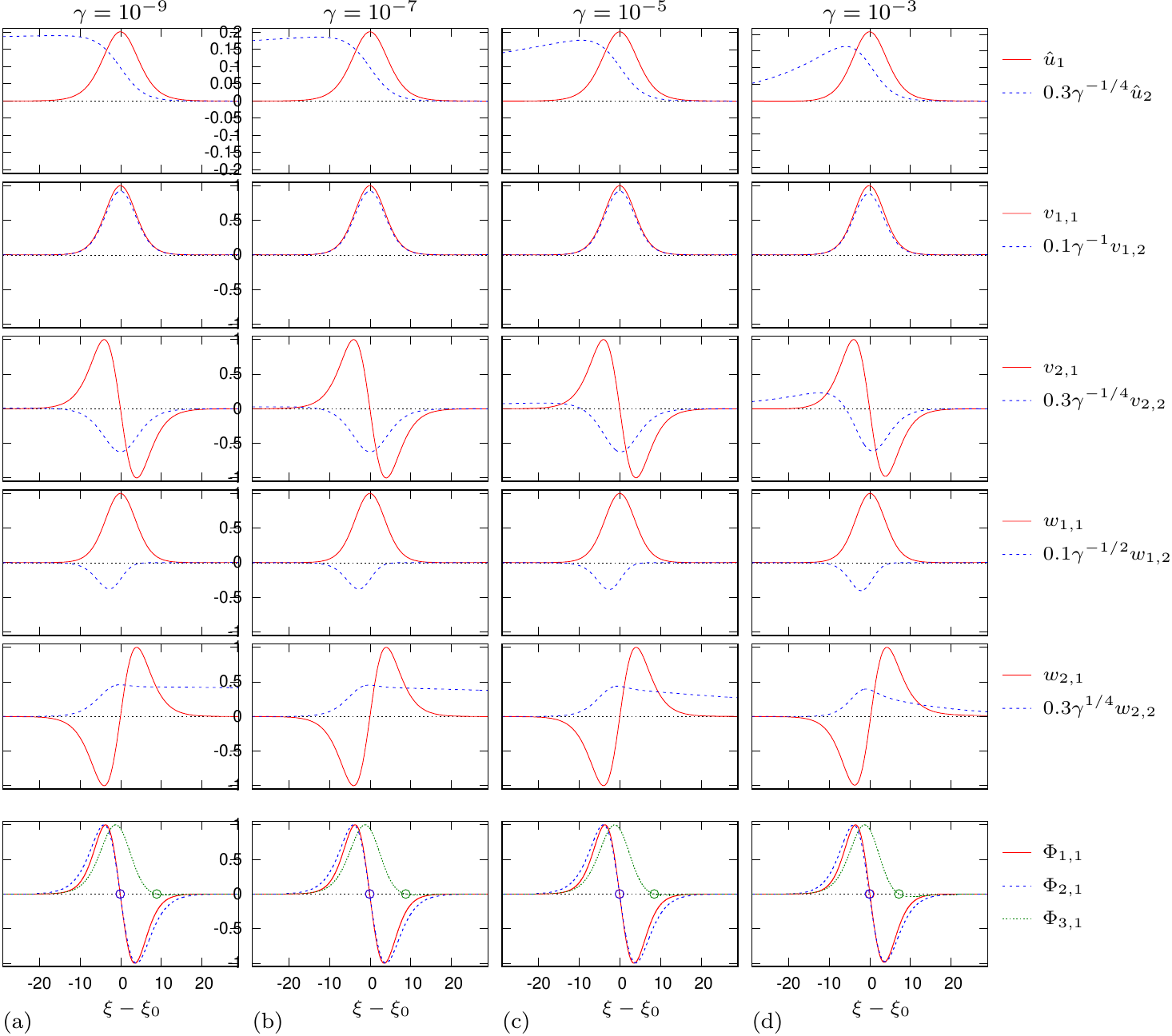}}{%
  Scaled ingredients of the linearized theory for the FHNCR
  model~\eq{fhnc} with varying $\gam$ as specified above the
  columns. Notations are the same as in \fig{fhn-sc}. %
}{fnc-sc}

Due to the degenerate dependence of $\fc1$ on $\ui2$, this model has
different asymptotic properties, discussed in
\subsecn{asymp-degenerate}. As expected from the results of that
subsection, the pulse speed scales as $\gam^{3/4}$, giving relatively
fast convergence of the pulse to a static nucleus state, see
\fig{fnc-nlin}(b).  Meanwhile, $\ui2 \sim \gam^{1/4}$, see
\fig{fnc-sc} (top row) --- a much slower convergence than standard FHN
($\q=1$).  Recall the pulse solution $\us$ asymptotically converges to
the critical nucleus solution $\u=[\nuc,0]$ as $\gam\to 0$.
Similarly, slower convergence is observed for components $\vi21$,
$\wi21$, and the component $\wi11$ asymptotically vanishes unlike its
counterpart in FHN model.  This is all in agreement with the
predictions from \subsecn{asymp-degenerate}. We are of course mostly
interested in how this difference affects the accuracy of the lineary
theory predictions for the critical curves.

\dblfigure{\includegraphics{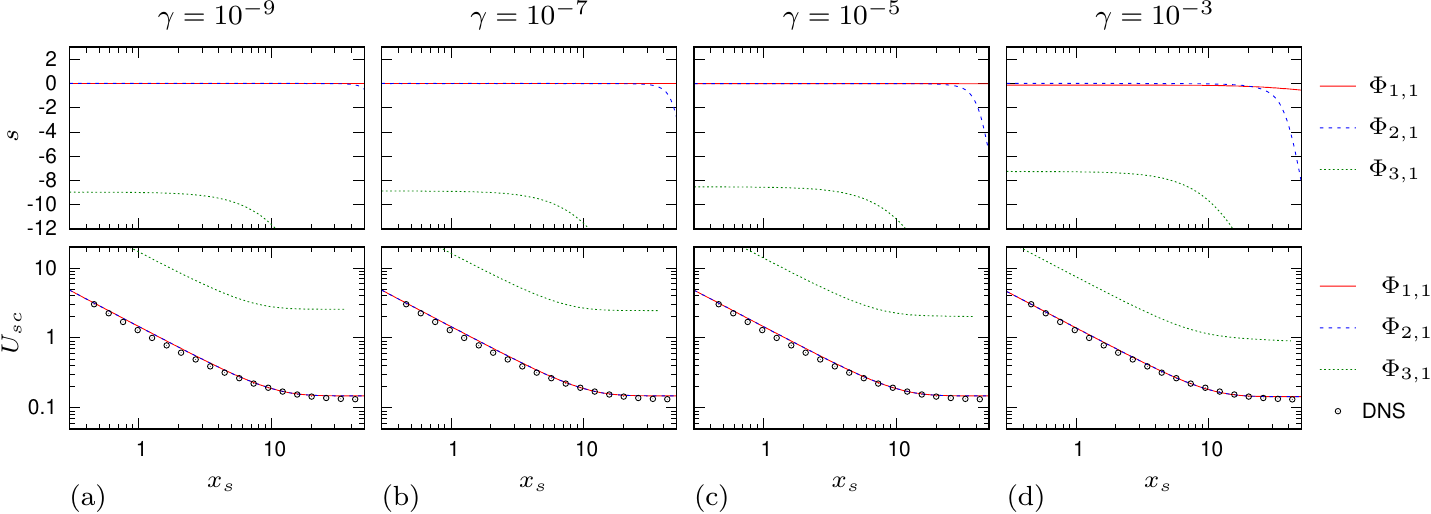}}{%
  Shifts $\shift$ predicted by the three different shift
  selectors $\Cmp\l$ (upper row), and the corresponding predicted
  critical curves against the direct numerical simulations (lower
  row), for the FHNCR model~\eq{fhnc}. %
}{fnc-ccs}

The FHNCR model results should be read in dialogue with the FHN
($\q=1$) model results.  As a point of concrete comparison, consider
$\gam=10^{-8}$ and $\xs \gg 10$ in the FHN model, \fig{fhn-ccs}, where
the frame selector $\Cmp3$ predicts an amplitude which is
approximately twice as large as the DNS result.  Compare to the same
configuration for the FHNCR model results for $\gam=10^{-7}$ and
$\gam=10^{-9}$, \fig{fnc-ccs}, which predicts a critical amplitude
which is an order of magnitude larger than the DNS result, and five
times larger than the linear model prediction.  While the frame shifts
selected by the $\Cmp3$ condition prove significantly worse for FHNCR
than for linear FHN, the frames selected by $\Cmp1$ and $\Cmp2$ are
perfectly adequate predictors for the critical amplitude for both
models.  That is, the predictive power of the linear theory is not
affected by the relative scaling of the components so long as a frame
selecting heuristic is chosen carefully.

Further, numerical experiments with the FHNCR (not shown) suggest
that, as $\q$ increases further, we should expect the predictive power
of $\Cmp1$ and $\Cmp2$ to outstrip that of $\Cmp3$ more generally.
Some slow-fast models of cardiac excitation, in particular, form a
highly nonlinear (and indeed, parametric) dependence of $\fc1$ on
$\ui2$.  We consider one such model in the next subsection.

\subsection{Karma}

The Karma-1994~\cite{Karma-1994} model is a qualitative model of
cardiac excitation, designed to reproduce chosen restitution curves.
The Karma model kinetics are given by the following functions,
\begin{align}\eqlabel{karma}
  \fc1 &= ((\kus - \ui2^\kM) (1 - \tanh(\ui1-\kud))\ui1^2/2 - \ui1) , \nonumber\\
  \fc2 &= \kbet \, \Heav_\k(\ui1-\kua) - \ui2 ,
\end{align}
where we will consider the parameter set $\kus = 1.5415$, $\kM=4$,
$\kud = 3.0$, $\kbet = 1.389$, $\kua = 0.5$, $\ktone = 2.5\,\ms$,
$\kttwo = 250\,\ms$, and in terms of the original notations
of~\cite{Karma-1994} we define $\gam \equiv \ktone/\kttwo$ as the ratio
of timescales, so that $\t$ is dimensionless.  The original diffusion
coefficient of the model is dimensional, we absorb this quantity in
the non-dimensionalization of the spatial scale, $\x$.

\sglfigure{\includegraphics{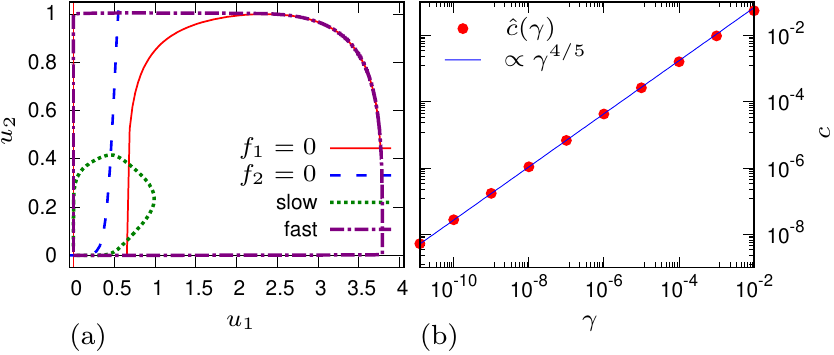}}{%
  Propagating waves in the Karma model~\eq{karma}.  (a) Phase plane of
  the point system with the null-clines and trajectories corresponding
  to the stable and unstable pulses, for $\gam=10^{-3}$. (b) Speed of
  the critical (slow) pulse as function of $\gam$, together with the
  theoretical asymptotic. %
}{k-nlin}

\Fig{k-nlin}(a) sketches a phase portrait of the kinetics and
\fig{k-nlin}(b) shows the scaling of the speed of the unstable pulse
solution for the Karma model.  The pulse speed scales as $\gam^{4/5}$,
which matches the predictions of~\subsecn{asymp-degenerate}; we of
course have $\q=\kM$ here.

\dblfigure{\includegraphics{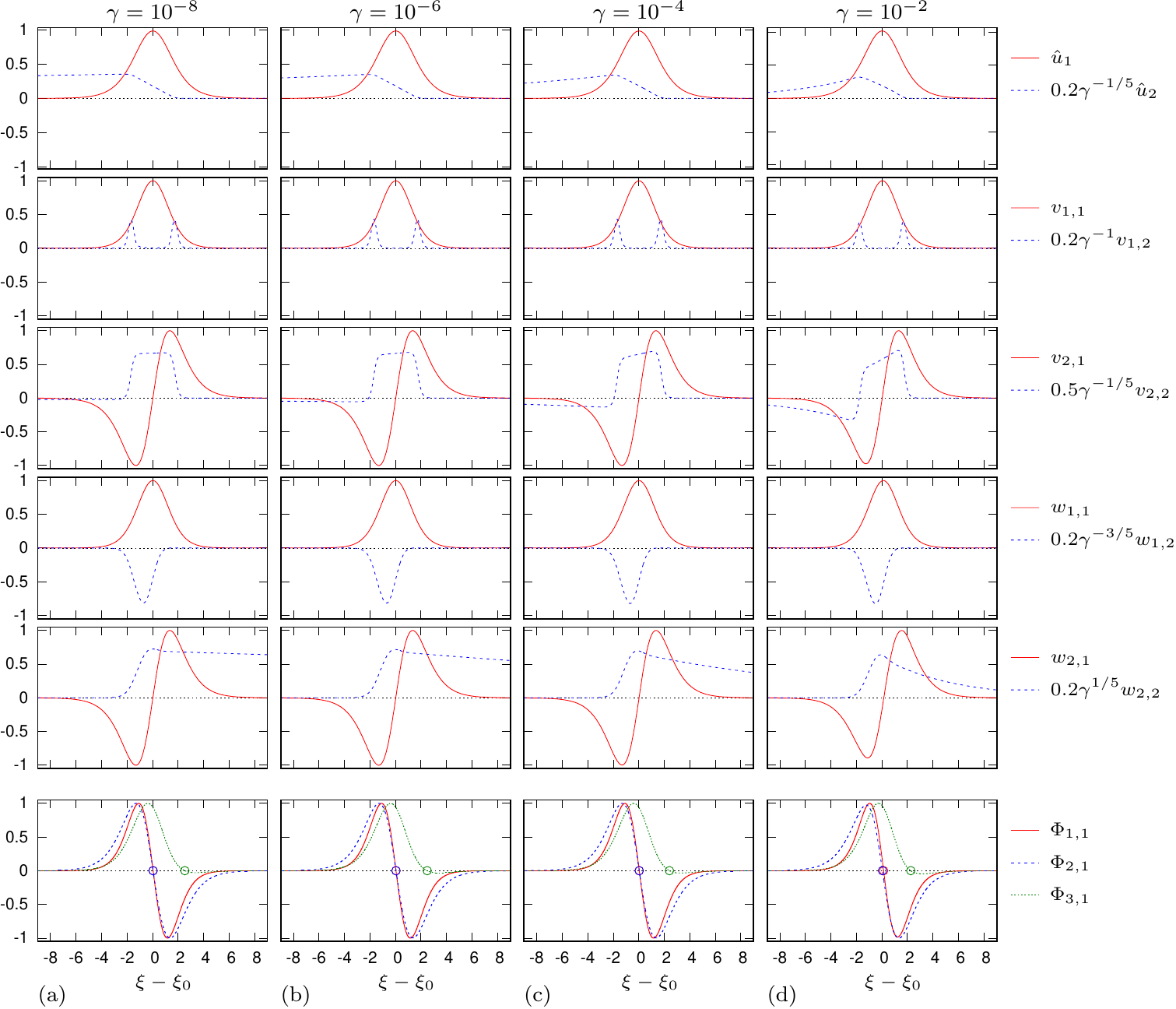}}{%
  Scaled ingredients of the linearized theory for the Karma model with
  varying $\gam$ as specified above the columns. The coordinate
  $\xf=\xo$ corresponds to the position of the maximum of $\usi1$.
  The thin vertical lines on the panels in the bottom row indicate
  position of the zeros of the corresponding shift selectors. %
}{k-sc}

\Fig{k-sc} summarizes the unstable pulse solution and leading
eigenfunctions.  As for the other models discussed above, the
amplitude of $\ui2$ decreases as $\gam\to 0$; however, for Karma the
convergence is significantly slower than the FHN or MS convergence
rates.  The convergence of the slow components of the leading left
eigenfunctions is likewise complicated by the quartic dependence on
$\ui2$.  In particular, the convergence rate for $\wi12$ and for
$\wi22$ are markedly different, suggesting that for some intermediate
value of $\gam$ the dominant contribution to $\cmp31$ switches from
the term proportional to $\wi21$ to the term proportional to $\wi22$,
and that predictions made on one side of the scale of $\gam$ will
contradict predictions made on the other, or asymptotically.  A simple
calculation suggests that this occurs for very large $\gam$, when the
asymptotic argument no longer holds.

While it is generically true that the structure of the leading
eigenfunctions of the unstable pulse changes as $\gam\to 0$, for the
generic models considered previously the localization of the
eigenfunction components are convergent for sufficiently small $\gam$.
For the FHNCR and Karma models, the leading right eigenfunctions
($\efr1$, $\efr2$) and the leading left eigenfunction ($\efl1$) are
well-behaved in the limit of small $\gam$.  Both the leading right
eigenfunctions ($\efr1$, $\efr2$) are localized to the same region as
the pulse, i.e., the leading right eigenfunctions inherit their
localization from the nonlinear solution.  Likewise, $\efl1$ is
localized like $\efr1$ and the first components of these
eigenfunctions coincide in the limit of small $\gam$.  However, as
$\gam\to 0$, the slow component of $\efl2$, $\wi22$, behaves
qualitatively differently.  While for the generic models, $\wi22$
converges to a localized function which decays quickly outside of the
central region of the unstable pulse, for the degenerate models it
does not. As $\gam\to 0$, $\wi22$ delocalizes asymmetrically, such
that $\abs{\wi22(\xf)} > 0$ ahead of the excited region of $\ui1$,
while $\abs{\wi22(\xf)} \to 0$ behind the peak.

\dblfigure{\includegraphics{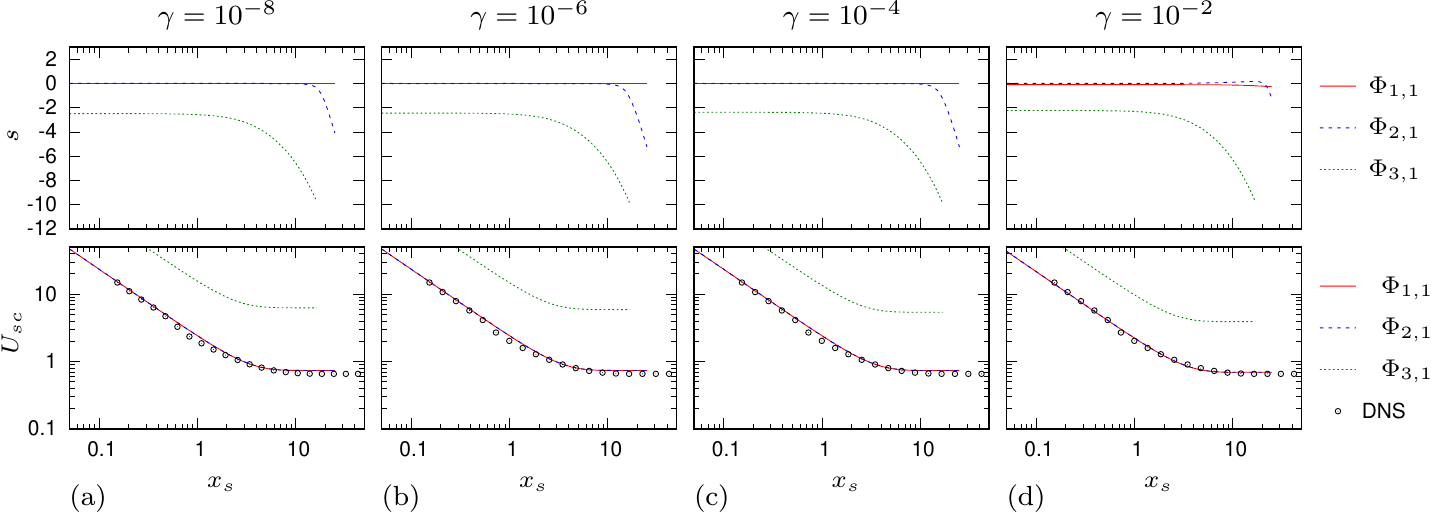}}{%
  Shifts $\shift$ predicted by the three different shift selectors
  $\Cmp\l$ (upper row), and the corresponding predicted critical
  curves against the direct numerical simulations (lower row), Karma
  model. %
}{k-ccs}

The Karma model results represent a stress-test for the asymptotic
scaling, both in terms of the stiffness of the kinetics and the
parametric nonlinear dependence of $\fc1$ on $\ui2$.  The former
specifically marks the asymptotic structure of the pulse as an
approximately piecewise linear curve, and the small-$\gam$ structure
of the leading right eigenfunctions as wildly deviant from the other,
smoother, models described in this work.  The latter presents an
opportunity to test the efficacy of the asymptotic analysis for severe
nonlinearity.

In addition to the quartic nonlinearity ($\kM=4$) results presented
here, we also computed the asymptotic scaling of the linear theory
ingredients for $\kM=2$, $\kM=8$, $\kM=16$, and $\kM=32$.  In each
instance of the parameterized model, the scaling of each component
follows the predicted asymptotics described
in~\subsecn{asymp-degenerate}.  As with the comparison of the cubic
and linear FHN model critical curve predictions, as the nonlinearity
$\kM$ increases, the less accurate $\Cmp3$ becomes, while $\Cmp1$ and
$\Cmp2$ maintain their predictive power across several decades of
$\xs$ and $\gam$.

\section{Discussion}

\paragraph{The mathematical problem addressed here} is that of
conditions required for initiation of a propagating excitation wave by
an instant perturbation from the resting state by a stimulus of a
certain spatial extent. The gist of our approach, previously exposed
in \cite{%
  Idris-Biktashev-2008,%
  Biktashev-Idris-2008,%
  Bezekci-etal-2015,Bezekci-2016,%
  Bezekci-Biktashev-2017,%
  Bezekci-Biktashev-2020%
}, is in linearization of the PDE system around a critical solution.
A delicate issue is translational invariance of the problem which
generates a one-parametric family of critical solutions, and poses a
problem of identification of the member of that family that
corresponds to a given initial perturbation. In the previous works,
this issue was addressed by an heuristic suggesting that the initial
condition of the linearized problem should not contain the shift
mode. In this framework, the essential ingredients of the linearized
theory, apart from the critical solution itself, are the right and
left eigenfunctions of the linearization, corresponding to the first
two eigenvalues, the first positive eigenvalue responsible for the
instability of the critical solution, and the second being the zero
eigenvalue corresponding to the translational symmetry. In particular,
the left eigenfunction corresponding to the zero eigenvalue serves as
the projector onto the shift mode.
 
In this paper, the focus is on four selected two-component systems
with slow-fast time scale separation, like in FitzHugh-Nagumo model
(FHN), including FHN model itself. In this class of models, the
critical solution is a slow unstable propagating pulse. The results
presented here were originally thought of as no more than further
tests of applicability of the above mentioned approach in a particular
class of models. This aim has been broadly achieved, however certain
unexpected aspects have been revealed, which may serve as valuable
lessons both for the specific problem but also at large for the theory
of slow-fast systems.

\paragraph{The expectation} was to verify an asymptotic
theory~\cite{Bezekci-Biktashev-2017} based on the would-be obvious
assumption that in the asymptotic limit, the events in the fast
subsystem dominate, and therefore the predictions of the linearized
theory should converge to those for the one-component system, in which
the slow variable is frozen at the resting state value. Specifically,
it was expected that the critical pulse solution converge to the
stationary unstable non-uniform ``critical nucleus'' solution of the
fast subsystem, and the eigenfunctions correspondingly converge to
those of the critical nucleus as far as the fast components are
concerned, and the slow components be negligible in the asymptotic
limit.

\paragraph{Lesson one: this expectation has proved wrong.} Although
the critical pulse solution does indeed converge as expected, and the
fast components of the eigenfunctions indeed converge as expected, but
the slow component paradoxically does not become
negligible. Specifically, the slow component of the second left
eigenfunctions, which is the projector to the translational mode and
is therefore material for the heuristic used for critical pulse
selection, grows large in the slow-fast asymptotic limit. This
eigenfunction corresponds to sensitivity of the speed of the
propagating pulse to perturbations of the slow component ahead of it,
and increases and spreads out in sync with slow-down of that
component. As a result, the overall contribution of the slow component
in the overlap integral, coming from the left and right eigenfunction,
does not vanish in the asymptotic limit, and although the predictions
based on ignoring the slow components work reasonably well, as it
appeared in the analysis done in~\cite{Bezekci-Biktashev-2017}, the
non-vanishing contribution from the slow component creates a
systematic error in identifying the critical pulse, which considerably
spoils prediction and in some cases renders the theory inappicable in
principle.

\paragraph{Lesson two: there are other heuristics which do withstand
  the asymptotic limit.} Heuristics are required in our approach
because of its leading idea to use linearization in the situation
where the solution in question is in fact not small. The old
heuristic, which has been in use since~\cite{Biktashev-Idris-2008},
was to make the linearization ``more applicable'', by making sure that
its initial condition is ``as small as possible'' in the sense that in
the critical situation, i.e. at the margin between successful and
unsuccessful initiation, not only the first generalized Fourier
component vanishes, which is a condition of criticality, but the
second component vanishes, too, which is always achievable by an
appropriate translation of the critial solution with respect to the
initiation stimulus. There are of course may other senses in which the
initial condition can be made ``as small as possible'', offering
alternative heuristics. In here we have explored two of them, one that
minimizes the initial condition of the linearized problem in $\Ltwo$
norm, and the other that minimizes the predicted threshold given by
the criticality condition. Both new heuristics do not depend on the
second left eigenfunction, and both have shown expectable convergence
in the asymptotic limit, i.e. to the critical nucleus results, and
good predictive ability, i.e. correspondence with the direct numerical
simulations.

\paragraph{Lesson three: asymptotics may not give good predictions
  even when the asymptotic parameter, the ratio of time scales, is indeed very small.} This of course can happen
if this is not the only small parameter in the
problem, but there are others, such as other time scale ratios, or
sharpness of transitions in the reaction kinetics, as we have seen in
the examples of Mitchell-Schaeffer and Karma-1994 models. In such
cases, the range of applicability of the asymptotics depends on those
other parameters, and may be well away from realistic parameter
ranges.

\paragraph{Lesson four: not all asymptotics are the same.} The natural
genericity assumptions about the dependence of the kinetics terms on
the dynamic variables may fail, leading to completely different
asymptotics in the slow-fast limits. A priori this possibility might
seem remote, but the fact that we have stumbled on such a failure ``by
accident'' in a popular, even if simplified, cardiac excitation
model~\cite{Karma-1994} suggests that this possibility should be kept
in mind. The slow rate of convergence in the fast/slow time scale
separation parameter, particularly together with other small
parameters present in the model, may render the fast/slow asymptotics
irrelevant, in the sense that the behaviour of the solution at the
original parameter value may be rather far from the asymptotic one,
even though the original parameter value appear rather small.

\paragraph{Further directions.} Straightforward extension of this
study would be to slow-fast systems with more than one fast and/or
more than one slow component with similar, Tikhonov type occurrence of
the small parameter.
Note that even within this paradigm, there may be
  qualitatively different types of excitability~\cite{Wieczorek-etal-2011,
Hesse-etal-2017}. A still
more intriguing possibility is about
asymptotics in case of non-Tikhonov slow-fast models, of the kind
discussed in~\cite{%
  Biktashev-Suckley-2004,%
  Biktashev-etal-2008,%
  Simitev-Biktashev-2011%
}. The known difference of systems with non-Tikhonov structure is that the
asymptotic limit of the critical solution is not a stationary
``nucleus'', but a moving front. Applicability of the linearized
theory has been tested on a conceptual model of such critical front
in~\cite{%
  Biktashev-Idris-2008,%
  Bezekci-etal-2015%
}. However, convergence of the ingredients of the linearized theory
and of corresponding predictions for the critical curves has not been
explored so far to our knowledge. Given the lessons from the
FitzHugh-Nagumo type systems discussed above, one should not take such
convergence for granted. This direction is particularly important
since non-Tikhonov asymptotics have been argued to better represent
the properties of realistic ionic models of cardiac excitation, than
FitzHugh-Nagumo type systems, particularly at the margins of
propagation.

Implications of the observations presented here are also relevant for
exotic solutions of one dimensional excitable models.  We have noted
that the convergence of the slow component of $\efl2$ is different for
the generic and degenerate models, not only in the scaling of the
solution near the asymptotic, but the asymptotic shape itself for
small $\gam$.  To reiterate, $\wi22$ is asymmetrically extended for
$x$ ahead of the pulse peak, while decays to zero quickly behind the
pulse peak.  This feature of $\efl2$ in conjunction with the
localization of $\efl1$ suggests that the slow dynamics (the
physically relevant dynamics) of the unstable pulse are nearly
insensitive to perturbations positioned post-peak, but very sensitive
to being slowed by perturbations almost arbitrarily ahead of the pulse
peak.  This may play an important role in the development of
``back-initiation'', or the observability of a ``one-dimensional
spiral'' solution generally~\cite{Cytrynbaum-Lewis-2008}.  The extension of
$\wi22$ should lead to acceleration of newly created pulse formations
ignited by back-initation and increase the potential for local
collapse to the resting state.  As we know such an unstable solution
exists in FHN-type models, one would expect that the increased
sensitivity of an extended $\wi22$ may suppress the formation of these
dynamics, suggesting that degenerate models may have more complex
saddle structures.

One natural extension of the existing program is to the ``critical
quenching problem'', that is, of cessation of stable wave propagation
in an excitable medium by addition of minimally invasive
perturbations.  The application to quenching is an inversion of the
application to ignition, though the central ingredients can be the
same and rely on the same linearization about the unstable pulse
state.  Crucially, as quenching considers an equivariant state in the
form of the stable wave, as compared to the invariant quiescent state,
the problem involves the consideration of an additional parameter
which fixes the additional translational symmetry.  This problem will
be addressed in a forthcoming paper.
 
\begin{acknowledgments}
  The authors thank Prof. Peter Ashwin for productive discussions
  throughout the creation of this manuscript. %
  This research was supported in part the EPSRC Grant No.
  EP/N014391/1 (UK), % CPMH
  and National Science Foundation Grant No. NSF PHY-1748958, NIH Grant
  No. R25GM067110, and the Gordon and Betty Moore Foundation Grant
  No. 2919.01 (USA). % Kavli
\end{acknowledgments}

\bibliographystyle{unsrt}

\end{document}